\documentclass[letterpaper, 10 pt, conference]{ieeeconf}  
                              
\usepackage{amsmath}
\usepackage{graphicx}
\usepackage{algorithmic}
\usepackage{algorithm}
\usepackage{multirow}

\IEEEoverridecommandlockouts                              
\title{\LARGE \bf
Answer Extraction in Question Answering using Structure Features and Dependency Principles
}

\author{Lokesh Kumar Sharma$^{1}$ and Namita Mittal$^{2}$
\thanks{*This work was not supported by any organization}
\thanks{$^{1}$L. K. Sharma is with Dept of Computer Science and Engineering,
        Galgotias College of Engg and Tech, Knowledge Park- II, Greater Noida, UP, IN
        {\tt\small lokesh at galgotiacollege.edu}}%
\thanks{$^{2}$N. Mittal is with Dept of Computer Science and Engineering, MNIT Jaipur, RJ, IN
        {\tt\small nmittal.cse at mnit.ac.in}}%
}

\begin{document}

\maketitle
\thispagestyle{empty}
\pagestyle{empty}

\begin{abstract}

Question Answering (QA) research is a significant and challenging task in Natural Language Processing. QA aims to extract an exact answer from a relevant text snippet or a document. The motivation behind QA research is the need of user who is using state-of-the-art search engines. The user expects an exact answer rather than a list of documents that probably contain the answer. In this paper, for a successful answer extraction from relevant documents several efficient features and relations are required to extract. The features include various lexical, syntactic, semantic and structural features. The proposed structural features are extracted from the dependency features of the question and supported document. Experimental results show that structural features improve the accuracy of answer extraction when combined with the basic features and designed using dependency principles. Proposed structural features use new design principles which extract the long distance relations. This addition is a possible reason behind the improvement in overall answer extraction accuracy.

\end{abstract}

	\section{Introduction}
	
Feature extraction (Agarwal et al. 2016) for question answering is a challenging task (Sharma et al. 2017). There are several answer extraction approaches (Severyn et al. 2013; Wei et al. 2006) which use the feature extraction. The issue with the existing techniques is that they work with limited features, and their success depends on a particular dataset. In this work, This issue has been resolved by proposing new features (i.e. Structural features), and this has been tested on variously available datasets (TREC and WebQuestions) and an original KBC dataset. For this firstly the features have been collected automatically (Yao et al. 2013) from an unstructured text document or a question. Few algorithms are designed to extract basic features using some design principles. The feature extraction algorithms are designed to extract new features from dependency parse of the question and the document. Prominent features are selected using feature selection techniques, and their relevance is decided using feature relevance techniques. In question answering task (Bishop et al. 1990; Brill et al. 2002; Bunescu et al. 2010), in vector space model, a question (Q) is represented as (Equation \ref{eq:31}):
	\begin{equation}
	Q = (f_1, v_1) , (f_2, v_2) , ..., (f_N, v_N)
	\label{eq:31}
	\end{equation}
	Where, $(f_i, v_i)$ is defined as \textit{$i^{th}$} feature and value of the question Q whereas, N $\in$ total number of features in Q. Due to the size of vector space particularly non-zero valued features are kept in the vector model. Therefore the size of individual features is pretty small despite the large size of feature space. These features are categorized into i) Basic features, and ii) Proposed features. Feature extraction algorithms are designed for both basic and proposed features. The basic features including all the lexical features, semantic features, and syntactic features are added to feature space.
	
\begin{figure}[ht!]
	\centering
	\includegraphics[width=0.48\textwidth]{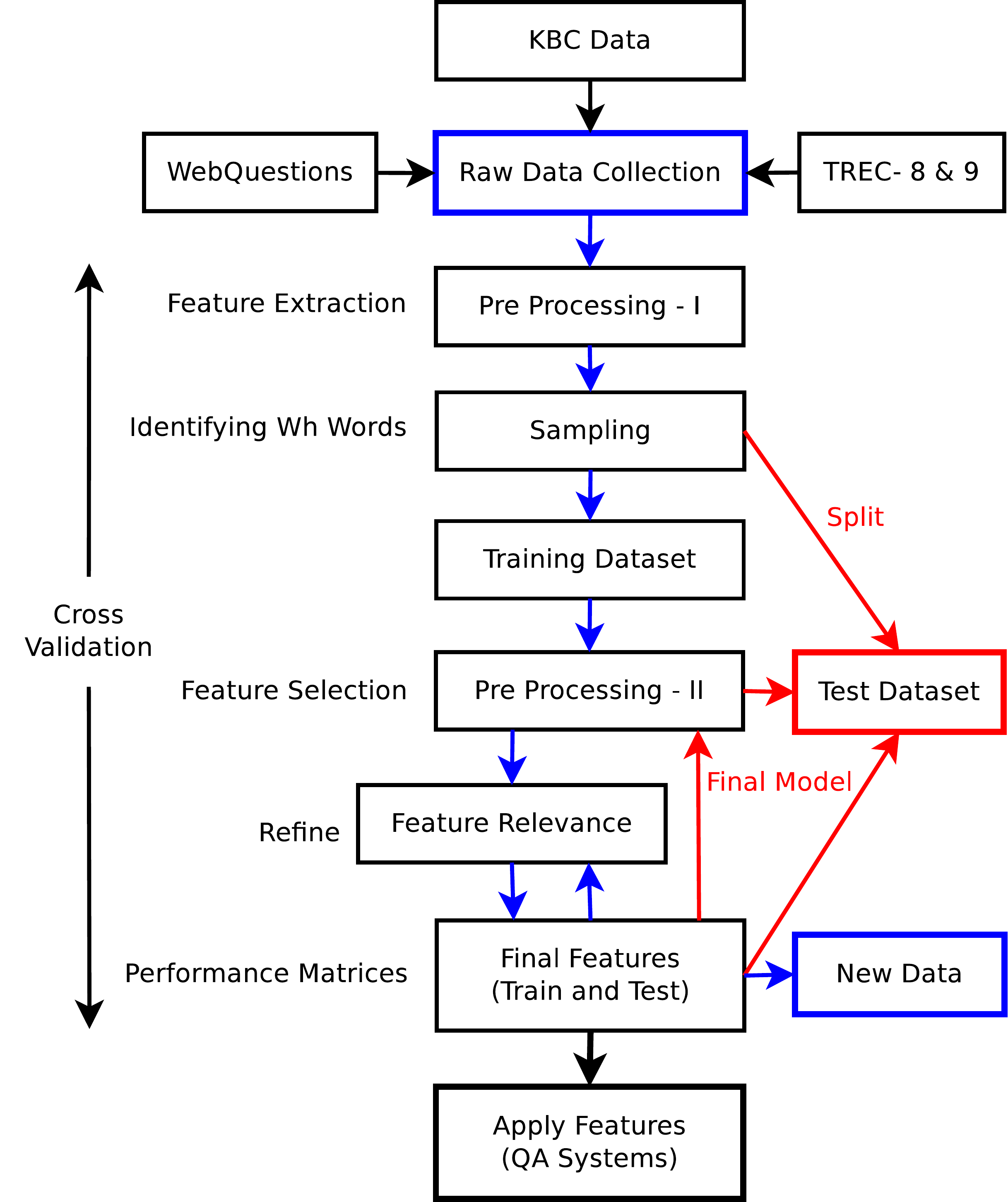}
	\caption{Raw dataset collection, features extraction, selection and new data generation for QA systems}\label{fig:31}
\end{figure}

The origin of these features and their extraction and selection procedure to create new dataset is shown in Figure \ref{fig:31}. The Figure \ref{fig:31} shows that the data is taken from the KBC game show questions of a particular episode (especially season-5). Apart from the KBC, the TREC (8 and 9) (Voorhees 1999; Singhal et al. 2000; Chowdhury 2003) and WebQuestions (WQ) (Wang et al. 2014) datasets are also selected. In the first stage, preprocessing is done, and features are extracted using feature extraction algorithms, and after the sampling process the dataset it is split into training and test question dataset. These datasets are further processed to select the relevant features and scaling is performed on these features. After this, relevant features are selected for training and testing to produce the final model. These features are applied for a successful answer extraction in QA. In the next sections, the two categories of features are discussed in details.

\section{Basic Features}

\begin{figure}[ht!]
	\centering
	\includegraphics[width=0.48\textwidth]{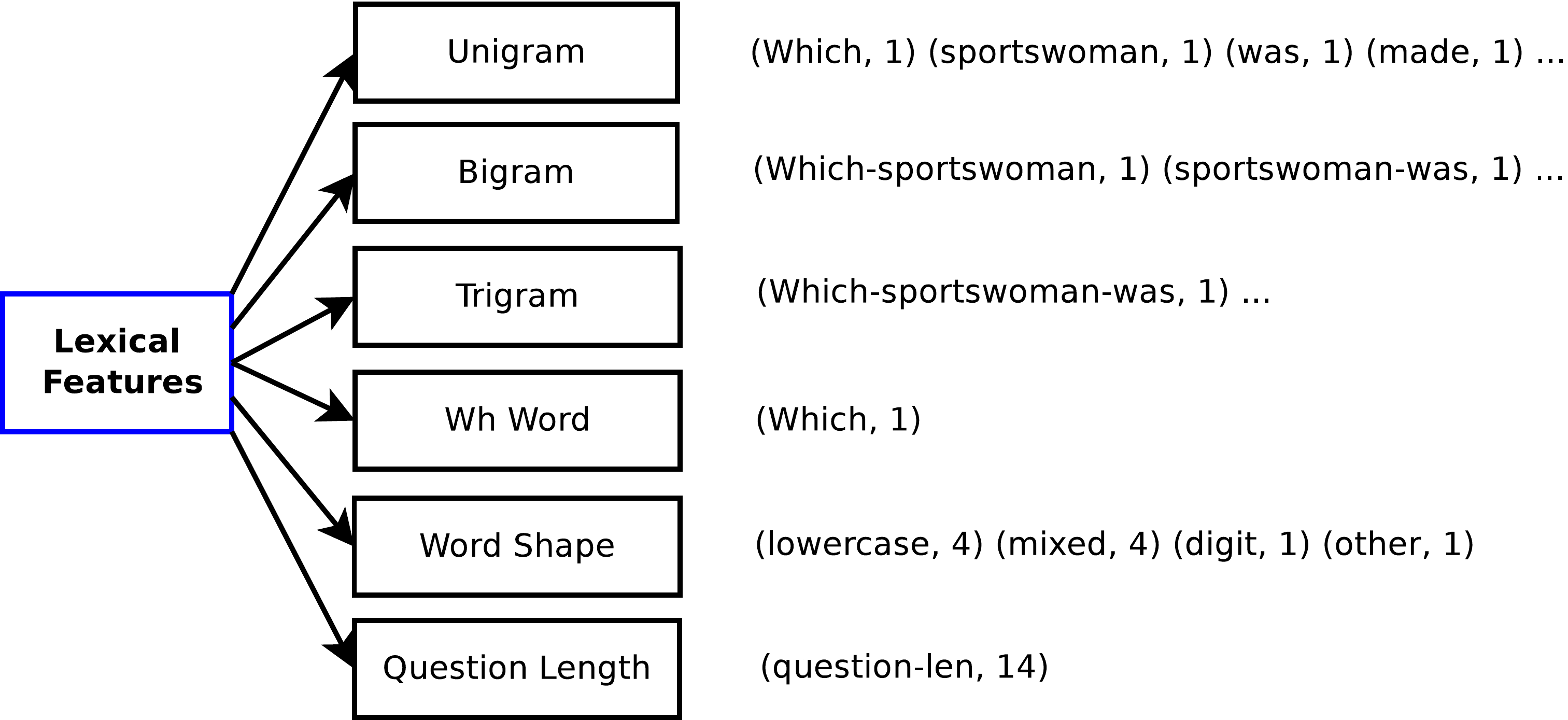}
	\caption{Lexical features present in a KBC question}\label{fig:32}
\end{figure}

\textbf{Lexical Features-} These are usually selected based on the words presented in the question. If we consider the single word as features is called unigram feature. Unigram is a particular case of the n-gram features. To extract n-gram features, a sequence of n-words in a question is counted as a feature. Consider for example the question \textit{`Which sportsperson was made the brand ambassador of the newly formed state of Telangana?'} from KBC dataset. Basic features of the lexical category are shown in Figure \ref{fig:32}.

Feature space for unigram is: $Q$ = (Which, 1), (sportsperson, 1), (was, 1), (made, 1), (the, 1), (brand, 1), (ambassador, 1), (of, 1), (newly, 1), (formed, 1), (state, 1), (of, 1), (Telangana, 1), (?, 1). The pair is in the form \textit{(feature, value)}, only the features with non-zero values are kept in the feature vector. The frequency of the words in question \textit{(feature values)} can be viewed as a weight value. It utilized this aspect to weight the features based on their importance. They joined different feature spaces with different weights. In their approach, the weight value of a feature space and the feature values (term frequencies) are multiplied. If any two consecutive words are considered as a different feature, then the feature space is extremely larger compared to unigram feature space and that demands larger training size. Therefore with same training set, unigrams perform better than bigrams or trigrams. In most of our experiments for answer extraction bigrams give better results than unigrams or other features.

Huang et al. (2011) examine a separate feature that is question's wh-words. They modified wh-words, namely which, how, where, what, why, when, who and remaining. For example, this feature of the question \textit{`What is the deepest ocean of the world?'} is \textit{`what'}. Considering the wh-words as a separate feature improves the performance of QA according to the experimental studies. The other kind of lexical feature is Word Shapes ($W_s$). It refers to possible shapes of the word: upper case, all digit, lower case, and other. Using word shapes alone is not a reliable feature set for question answering, but their combination with another feature improve the performance of QA.
The another lexical feature is question’s length; it is a total number of words in the question. The features are represented in a similar way to the Equation \ref{eq:31}.

\textbf{Syntactical Features-} The most basic syntactical features are Part of Speech (POS) tags and headwords. POS tags indicate such as NP (Noun Phrase), JJ (adjective), etc. The above mentioned the pos tags: \textit{Which/WDT sportsperson/NN was/VBD made/VBN the/DT brand/NN ambassador/NN of/IN newly/RB formed/VBN state/NN of/IN Telangana/NNP}. A POS tagger obtains the pos tags of a question. In QA, all the pos tags of a question in feature vector can be added applied as bag-of-pos tags.

Some more feature namely tagged unigram which is a unigram expanded with part-of-speech tags. Instead of using common unigrams, tagged unigrams can help to identify a word with different tags as two separate features. 

\begin{figure}[ht!]
	\centering
	\includegraphics[width=0.48\textwidth]{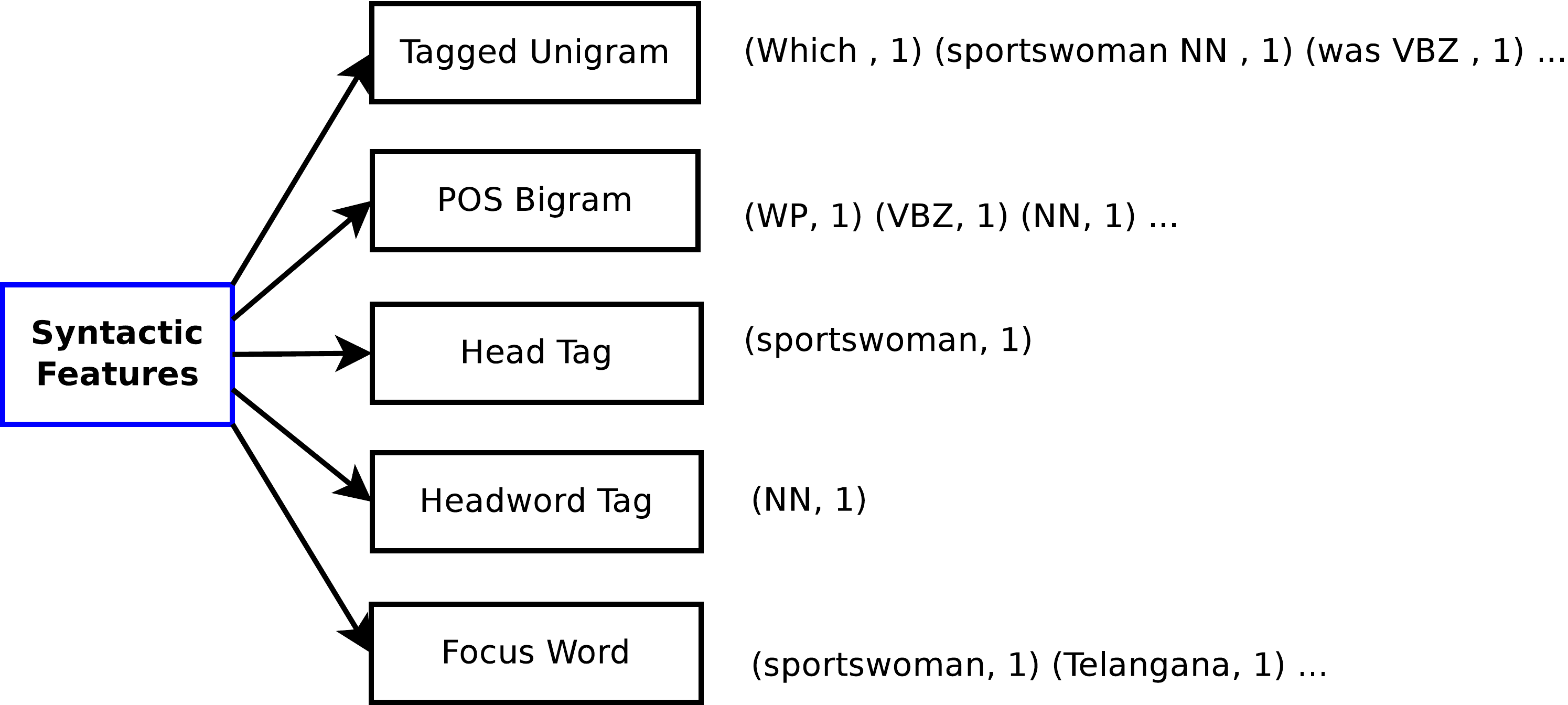}
	\caption{Syntactic features present in a KBC question}\label{fig:33}
\end{figure}
In syntactic features, headword is the most edifying word in a question or a word that represents the object that question attempts. Identifying a headword can improve the efficiency of a QA system. For example for the question \textit{`Which is the newly formed state of India?'}, \textit{`state'} is the headword. The word \textit{`state'} majorly contribute to classifier to tag \textit{LOC:state}. Extracting question’s headword is challenging. The headword of a question frequently selected based on the syntax tree of the question. To extract the headword, it is required to parse the question to form the syntax tree. The syntax (parse) tree is a tree that represents the syntactical structure of a sentence base on some grammar rules. Basic syntactic features are shown in Figure \ref{fig:33}.

\textbf{Semantic Features-} These are extracted from the question on the basis of the meaning of the words in a question. Semantic features (Corley et al. 2005; Islam et al. 2008; Jonathan et al. 2013) require third party resources such as WordNet (Miller 1995) to get the semantic knowledge of questions. The most commonly using semantic features are hypernyms, related words, and named entities.

Hypernyms are the lexical hierarchy with important semantic notions using the Wordnet. For example, a hypernym of the word \textit{`school'} is \textit{`college'} of which the hypernym is \textit{`university'} and so on. As hypernyms provide abstract over particular words, they can be useful features for QA. Extracting hypernyms is not easy as, 

\begin{enumerate}
	\item It is difficult to know the word(s) for which one need to find the hypernyms? 
	\item Which part-of-speech should be counted for focus word selection? 
	\item  The focus word(s) expanded may have several meanings in WordNet. Which meaning is to be used in the given question? 
	\item Which level can one go to the hypernym tree to achieve the prominent set?
\end{enumerate}
To overcome the problem of obtaining a proper focus word. The question can consider the headword as the focus word and it can be expanded for its hypernyms. All nouns in a question are considered as candidate words. If the focus word and the hypernym are same, this word can be expanded further. Consider the question again \textit{`What is the most populated city in India?'}. The headword of this question is \textit{`city'}. The hypernym features of the word with value six as the maximum depth will be as follow: {(area, 1) (seat, 1) (locality, 1) (city, 1) (region, 1) (location, 1)}. The word \textit{`location'} features, can contribute the classifier to categorize this question to \textit{LOC}.

\begin{figure}[ht!]
	\centering
	\includegraphics[width=0.48\textwidth]{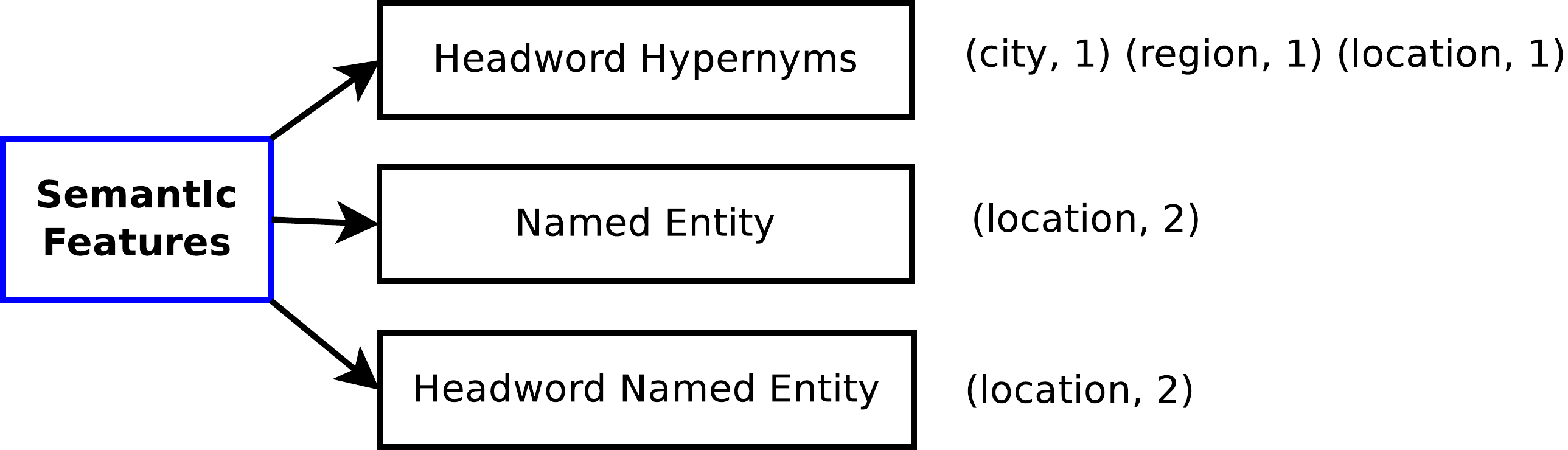}
	\caption{Semantic features in a KBC question}\label{fig:34}
\end{figure}

Named entities are the predefined categories of name, place, time etc. The available methods are applied to achieved an accuracy of more than 92.0\% on determining named entities. For example for the question, \textit{`Who was the second person to reach at the moon surface?'}, their NER system identifies the following named entities: \textit{`Who was the [number second] person to reach at the [location moon surface]?'} In question answering the identified named entities improves the performance when added to the feature vector. Basic features of a lexical category are shown in Figure \ref{fig:31}. Apart from these basic features, proposed features for answer extraction are discussed in the next section. Figure \ref{fig:34} shows the basic semantic features.

\section{Proposed Features and Feature Extraction Algorithms}
The proposed structural features in a question are extracted from its dependency parse (Lally et al. 2012; Pedoe et al. 2014) with additional Design Principles (DP), discussed later in details. These new features equally contribute in feature vector which is used for EG, NLQA and answer extraction in QA systems. Before going into details of the proposed features and their feature extraction rules, the feature extraction algorithms for basic features have been discussed. The steps of the algorithm is explained in the next sub section. The Algorithm 1 is used to extract all six lexical fatures to form a lexical feature vector of basic features.

\begin{algorithm}
 	\caption{\textbf{Algorithm 1:} Lexical feature extraction}
	 \textbf{INPUT:} Question set (Q) \\
	 \textbf{OUTPUT:} $Le_{fv}$ $\in$ Lexical feature vector from Q\\
	 \textbf{Variables Used:}\\
	     $(Q_f, V) \in (Question Feature, Feature Value)$ \\
	 $TF \in Term Frequency$  \\
	 $DL \in Document Length$  \\
	 $QL \in Question Length$ 
	\begin{algorithmic} [1]
	\FOR {questions in dataset $Q$} 
		\IF{$Q_w \neq `?'$}
		\STATE extract lexical features of the question
		\ENDIF
		\IF {$L_e$ extract a unigram}
		\STATE $(Q_{ui}, V_{ui}) \leftarrow Unigram_i$
		\STATE $V_{ui} = \frac{TF}{DL}$
		\STATE $V_{ui} \in Feature Value of i^{th} Unigram$
		\IF {$L_e$ extract a bigram}
		\STATE $(Q_{bi}, V_{bi}) \leftarrow Bigram_i$
		\STATE $V_{bi} = \frac{TF}{DL}$
		\STATE $V_{bi} \in Feature Value of i^{th} Bigram$
		\IF {$L_e$ extract a trigram}
		\STATE $(Q_{ti}, V_{ti}) \leftarrow Trigram_i$
		\STATE $V_{ti} = \frac{TF}{DL}$
		\STATE $V_{ti} \in Feature Value of i^{th} trigram$
		\ENDIF    
		\ENDIF
		\ENDIF
		\IF {input is $Q_i$ ($i^{th} question$)}
		\STATE $W_w \gets$ $WhWordList$
		\STATE $(Q^{wh}_i, V^{wh}_i) \leftarrow \sum WhWord_i$
		\STATE $(Q_{wsi}, V_{wsi}) \leftarrow WordShape$
		\STATE $V_{wsi} = WordShape$
		\STATE $V_{wsi} \in Feature Value of i^{th} WordShape$ 
		\IF {$QL \neq 0$}
		\STATE $QL \leftarrow QuestionLength$
		\ENDIF
		\ENDIF
		\ENDFOR
		\STATE \textbf{Return} $ \rightarrow LexicalFeatureVector$ $(Le_{fv})$ 
	\end{algorithmic}

\end{algorithm}

\subsection{Explanation of Algorithm to Extract Basic Features}
Lexical features are easy to extract because these are obtained from the question, and no third party software (e.g. WordNet) is required. For a given question set (e.g. KBC) all lexical features are extracted to form a feature vector called lexical feature vector ($Le_{fv})$. In Algorithm 1, from step 2 to 4 question length is checked. Than starting from step 5 to 8, step 9 to 12 and 13 to 16 the Unigram, Bigram and Trigram features are extracted respectively. These three features have been added in to the feature vector. Than from step 20 to 29 mainly the Word Shape and Question Length features are added in to the feature vector. A $ \rightarrow LexicalFeatureVector$ $(Le_{fv})$ is the outcome of this overall algorithm. This $(Le_{fv})$ is used to train the question model on the lexical features.

\begin{enumerate}
	\item $check\_Q$ $(Question\_Termination)$ (Lines 1 to 7 in Algorithm 1)
	\item $extract\_NG$ $(N\_grams)$ (Lines 8 to 22 in Algorithm 1)
	\item $extract\_WhWs$ $(Whword\_Wordshape)$ (Lines 24 to 29 in Algorithm 1)
	\item $extract\_QL$ $(Question\_Length)$ (Lines 30 to 32 in Algorithm 1)
	
\end{enumerate}
In Algorithm 1, the combined feature extraction algorithm for all lexical features shown. The accuracy of each feature is shown in Table \ref{table:31} in the end of this section. Total 500 KBC questions are used to examine the feature extraction accuracy, and the algorithm attains 100\% feature extraction accuracy for all lexical features. Now, syntactic features are extracted in the similar manner shown in Algorithm 2.

\begin{algorithm}
 	\caption{\textbf{Algorithm 2:} Syntactic feature extraction}
	 \textbf{INPUT:} Question set (Q) \\
	 \textbf{OUTPUT:} $Sy_{fv}$ $\in$ Syntactic feature vector from Q\\
	 \textbf{Variables Used:}\\
	     $(Q_f, V) \in (Question Feature, Feature Value)$ \\
	 $TF \in Term Frequency$  \\
	 $DL \in Document Length$  \\
	 $QL \in Question Length$ 
\begin{algorithmic} [1]
		\FOR {questions in dataset $Q$} 
		\IF{$Q_w \neq `?'$}
		\STATE extract syntactic features of the question
		\ENDIF
		\IF {$S_y$ extract a tagged\_unigram}
		\STATE $(Q^t_{ui}, V^t_{ui}) \leftarrow TaggedUnigram_i$
		\STATE $V^t_{ui} = \frac{TF^{tu}}{DL}$
		\STATE $V^t_{ui} \in Feature Value of i^{th} TaggedUnigram$
		\IF {$S_y$ extract a POS\_tags}
		\STATE $(Q_{pi}, V_{pi}) \leftarrow Stanford\_tagger$
		\IF {$S_y$ extract a Headword}
		\STATE $(Q_{hi}, V_{hi}) \leftarrow Headword$ (using headword extraction algorithm)
		\STATE $(Q^t_{hi}, V^t_{hi}) \leftarrow HeadwordTag$
		\STATE $(Q^h_{hi}, V^h_{hi}) \leftarrow Wordnet$ ($Q^h_{hi}$ $\in$ headword hypernym of i$^{th}$ word )
		\ENDIF    
		\ENDIF
		\ENDIF
		\IF {input has multiple $headword$)}
		\STATE $(Q^w_{fi}, V^w_{fi}) \leftarrow focusword_i$
		\ENDIF
		\ENDFOR
		\STATE \textbf{Return} $ \rightarrow SyntacticFeatureVector$ $(Sy_{fv})$ 
	\end{algorithmic}
\end{algorithm} 

\subsection{Explanation of Algorithm to Extract Syntactic Features}
For a given question set all syntactic features are extracted to form a feature vector called syntactic feature vector ($Sy_{fv})$. Apart from similarities in Algorithm 1 and Algorithm 2, from step 8 to 11 tagged unigram is extracted. Than from step 12 to 20 Headword and Headword tag is extracted. Later in step 21 it is checked for multiple headwords. A $ \rightarrow SyntacticFeatureVector$ $(Sy_{fv})$ is the outcome of this overall algorithm. This $(Sy_{fv})$ is used to train the question model on the syntactic features.

Syntactic features are quite difficult to extract because these features are extracted from the question and also require a third party software (e.g. WordNet). For a given question set (e.g. KBC) all syntactic features are extracted and placed into the features vector, it is called a syntactic feature vector ($Sy_{fv})$.

Algorithm 2 is showing the combined feature extraction algorithm for syntactic features, and the accuracy of each of these features is shown in Table \ref{table:31} in the end of this section. 

\begin{figure}[ht!]
	\centering
	\includegraphics[width=0.48\textwidth]{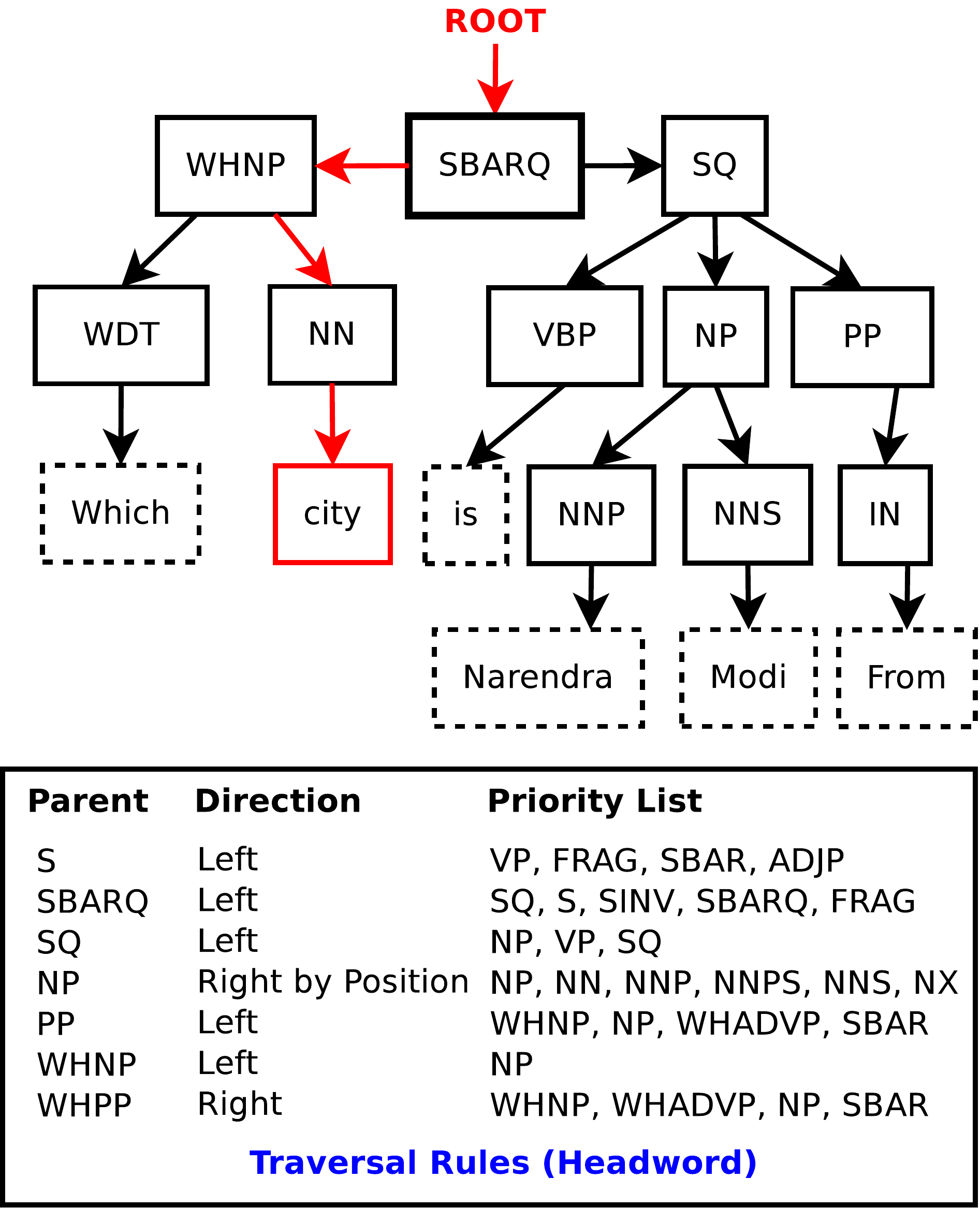}
	\caption{Tree traversal rules for headword}\label{fig:35}
\end{figure}

\begin{enumerate}
	\item $check\_Q$ $(Question\_Termination)$ (Lines 1 to 7 in Algorithm 2)
	\item $extract\_TP$ $(Taggedunigram\_Postag)$ (Lines 8 to 20 in Algorithm 2)
	\item $extract\_H$ $(Headword)$ (Lines 14 to 17 in Algorithm 2)
	\item $extract\_FW$ $(Focus\_Words)$ (Lines 21 to 23 in Algorithm 2)
\end{enumerate}
The tree traversal rules shown in Figure \ref{fig:35} are implemented to get a headword which is used as a significant syntactic feature of the question. The accuracy of this headword extraction algorithm is 94.3\% on KBC questions as the traverse rules are formulated manually.

\begin{table}
	\caption {\footnotesize{Accuracy of lexical feature extraction algorithms}}
	
	\begin{center}
		\scalebox{0.88}{
			\begin{tabular}{|l|c|c|c|}
				\hline
				\multicolumn{4}{|l|}{\textbf{Accuracy of Lexical Feature Extraction Algorithm}} \\ 
				\multicolumn{4}{|l|}{\textbf{Total No. of Question = 500}} \\ 
				\hline
				\textbf{Lexical Feature}   & \multicolumn{2}{l|}{\textbf{Features Extracted}}     &  \textbf{Accuracy} \\ \cline{2-3}
				& \textbf{Correct}    &    \textbf{Incorrect} &  \\ \hline
				Unigram ($U_n$) & 500    &    0 & 100\% \\ \hline
				Bigram ($B_i$) & 500    &    0 & 100\% \\ \hline
				Trigram ($T_r$) & 500    &    0 & 100\% \\ \hline
				Wh-Word ($W_w$) & 500    &    0 & 100\% \\ \hline
				Word Shape ($W_s$) & 500    &    0 & 100\% \\ \hline
				Question Length ($Q_l$) & 500    &    0 & 100\% \\ \hline
				
		\end{tabular}}
	\end{center}
	
	\label{table:31} 
\end{table}
The Table 1 is showing the accuracy of feature extraction on basic (lexical) features. The accuracy of finding the correct headword is 94.3\% (as discussed), it can be improved by learning methods.

The headword extraction algorithm in Algorithm 3 is using the traversal rules shown in Figure \ref{fig:35}. Loni (2011) uses these traversal rules for headword extraction for question classification. For example, for the question \textit{``Who was the first man to reach at the moon?"} the headword is \textit{``man"}. The word \textit{``man"} will contribute for getting the Expected Answer Type (EAT). The Algorithm 3 extracts the headword using the rulebase (Mohler et al. 2009).

\begin{algorithm}
	\caption{\textbf{Algorithm 3:} Headword extraction algorithm}
	\label{algo:33}
	
	\begin{algorithmic} [1]
		\STATE \textbf{procedure} Extract\_Tree
		\IF{$isTerminal(tree)$}
		\STATE \textbf{return} tree
		\ELSE
		\STATE root\_node $\leftarrow$ apply-traversal-rules (tree)
		\STATE \textbf{return} Extract-Question-Headword
		\ENDIF
		\STATE \textbf{end procedure}
		
	\end{algorithmic}
\end{algorithm}

Few examples are showing the headword of a question. The words in bold are the possible headwords. There should be clear mention for a headword and a focus word.

\begin{enumerate}
	\item[] What is the nation \textbf{flower} of India?
	\item[] What is the name of the \textbf{company} launched JIO 4G in 2016?
	\item[] What is the name of world's second longest \textbf{river}?
	\item[] Who was the first \textbf{man} to reach at the moon?
\end{enumerate}

\subsection{Proposed Structural Features}
The proposed structural features are obtained from the features in the yield of dependency parse. These structural features (say, $S_t$) are employed for complicated relations presented in similar questions and used for the uniqueness of efficient constants available in parsing results. 

\begin{figure}[ht!]
	\centering
	\includegraphics[width=0.48\textwidth]{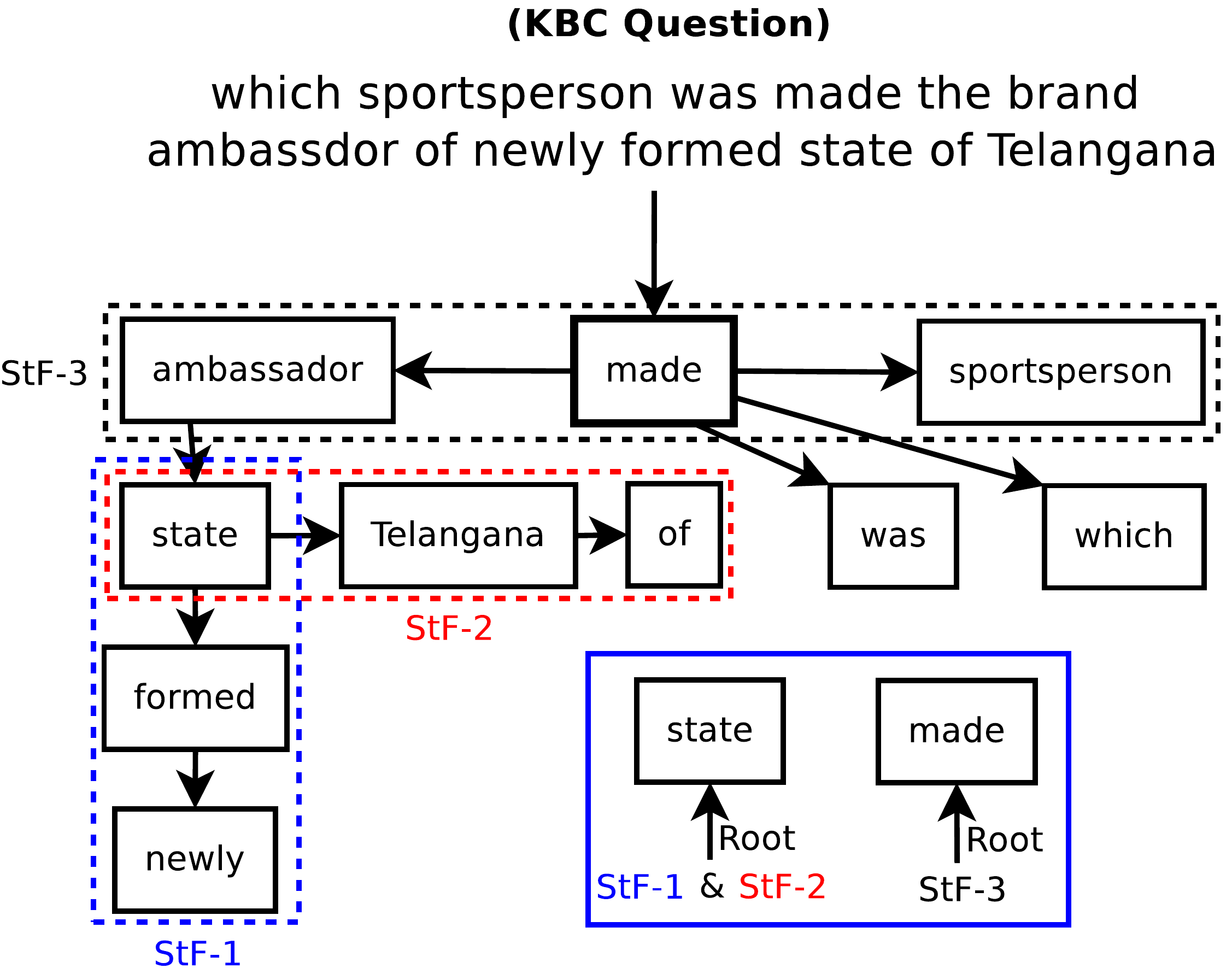}
	\caption{Structural features in a question used to align two words}\label{fig:36}
\end{figure}

The Question Feature Form (QFF) produced for a question contains one composite feature function. Structural features allow the model to adapt for all questions used for alignment using the question structure. Figure \ref{fig:36} is showing the structural features available in a KBC question. There are some relations where \textit{`state'} can be aligned with \textit{`newly formed'} and \textit{`of Telangana'} and another structural feature where \textit{`made'} is aligned with \textit{`ambassador'} and \textit{`sportsperson'}. The link between \textit{newly-formed $\rightarrow$ Telangana} and \textit{state $\rightarrow$ made} cannot be identified directly. The connection provides a structural confirmation which has been described in details later in this section.

\subsubsection{Dependencies Rules for Structural Features}
Researchers in different domains have successfully used dependency Rules (DR) or Textual Dependencies (TD). In Recognizing Textual Entailment the increase in the application TD is distinctly apparent. 

\begin{figure}[ht!]
	\centering
	\includegraphics[width=0.48\textwidth]{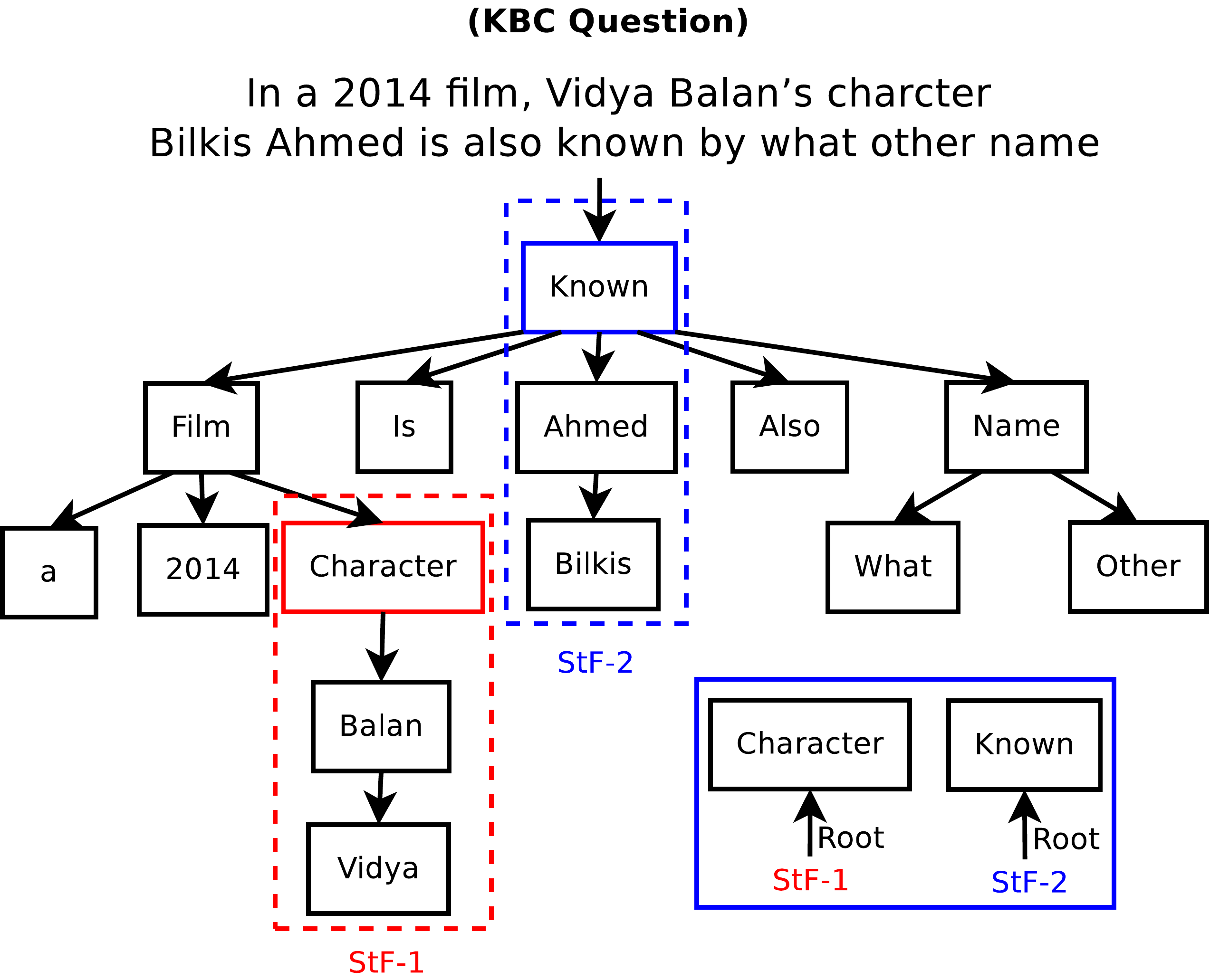}
	\caption{Structural features in a KBC question}\label{fig:37}
\end{figure}

It is found that the rules are designed from the dependencies in the extraction of a relation between question and document, a system with DR semantics considerably outperforms the previous basic features on KBC dataset (by a 9\% average improvement in F-measure). The tree-based approach uses a dependency path to form a graph of dependencies. The system those uses TD demonstrates improved performance for the feature-based techniques.

In the Figure \ref{fig:37} structural features of a KBC questions are highlighted. The parsing technique uses the relation \textit{`Vidhya Balan, a film character, has worked as Ahmed Bilkis in 2014'} separated by commas on the NP. The Parser uses a diverse dependency presented in questions and relevant document. Another example is the PP where many relations mean an alternative attachment with structures. By targeting semantic dependencies TD, provides an adequate representation for a question.

\begin{figure}[ht!]
	\centering
	\includegraphics[width=0.48\textwidth]{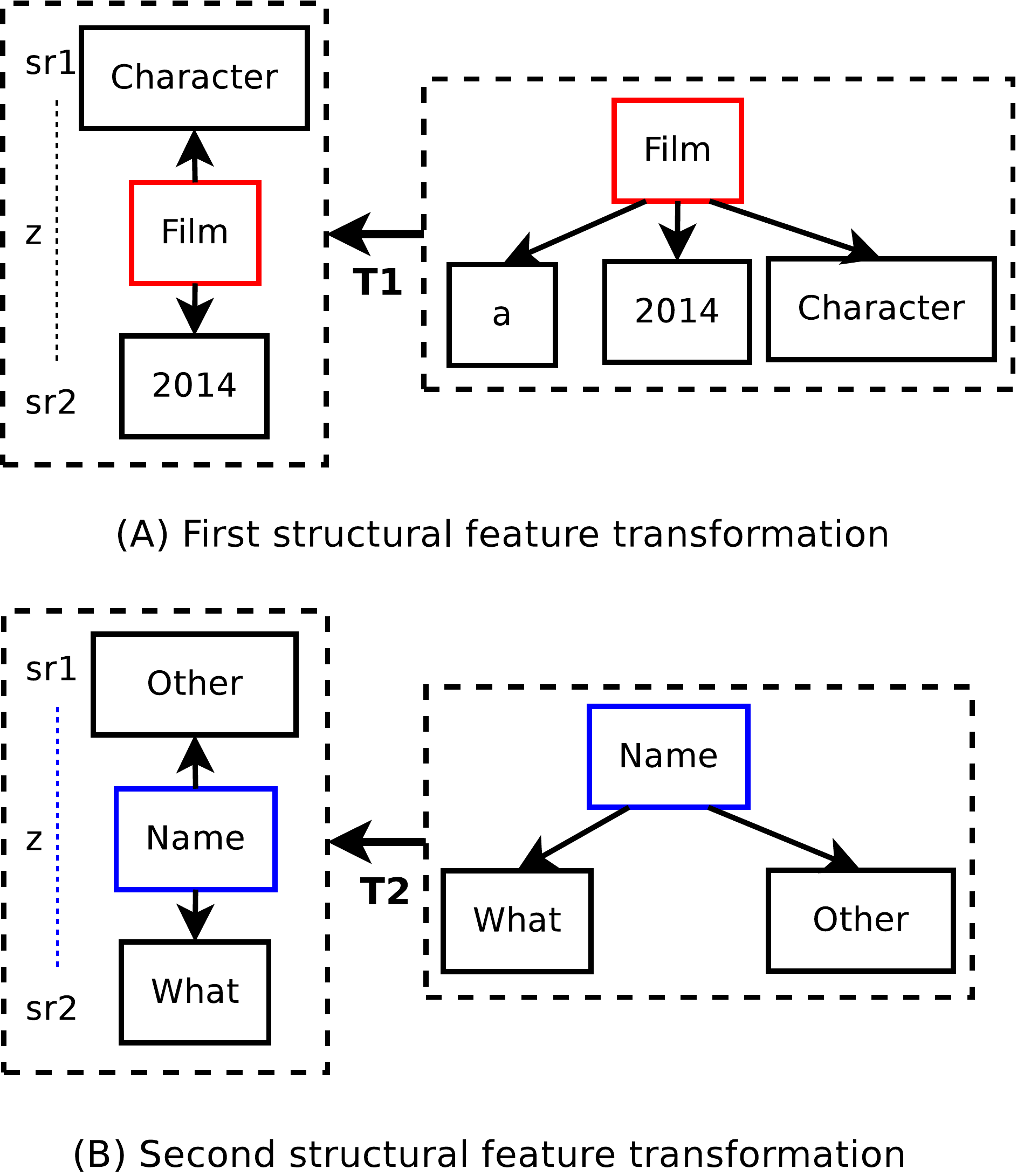}
	\caption{Transforming structural features into binary relations}\label{fig:38}
\end{figure}

The structural features are transformed into a binary relation by removing the non-contributing words (i.e. stopwords). Figure \ref{fig:38} show such a design for two structured features $T_1$ and $T_2$ of a question which is shown in Figure \ref{fig:38}. There can be more than two structural features in the question so there can be more than two structural transformations.

KBC dataset has manually annotated data for information retrieval in the open domain version to be tagged with the TD scheme. These conversion rules that are used to transform TD tree into a binary structural relation.

\subsubsection{Design Principals for structural features}
The structural feature representation bears a strong representation of feature vector space, and, more directly, it describes the grammatical relations (Severyn et al. 2013; Sharma et al. 2015). These design principals are used as a starting point for extracting the structural features. For obtaining SF, the TD helps in structural sentence representation, especially in relation extraction. SF makes available two options: in one, relations between it and other nodes, whereas in the second, making changes and adding prepositions into relations. 

The intended use of structural extraction SF attempt to adhere to these six design principles (DP$_1$ to DP$_6$):\\
\textbf{DP$_1$:} Every dependency is expressed as a binary relation obtained after the structural transformation. \\
\textbf{DP$_2$:} Dependencies should be meaningful and valuable to EG and NLQA. \\
\textbf{DP$_3$:} The structural relations should use concepts of traditional grammar to link the most frequently related word. \\
\textbf{DP$_4$:} The relations with a maximum number of branches should be available to deal with resolving the complexities of indirect relations helpful for alignment. \\
\textbf{DP$_5$:} There should be the maximum possibility of relations to be in NP words and should not be indirectly mentioned via non-contributing words. \\
\textbf{DP$_6$:} Initially the is the longest meaningful connection on which minimum non-contributing words than linguistically expressed relations. \\ From dependency rules and design principals for structural features the feature extraction algorithm aims to extract all possible structural features of the question. This structural feature extraction algorithm considering these design principles is discussed in the next section.

\subsubsection{Structural Feature Extraction Algorithm and Score}
\par The proposed structural features which are obtained from the dependency structure of a question on the basis of DPs. Consider $X$ to be the root for this relation ($X \rightarrow Y$) $Y_i$ will participate in structural features.

\par The relations of dependency tree (Wei et al. 2006) are used for extracting SFs to capture long-distance connections in the question and text. For the sentence: \textit{`Which sportsperson was made the brand ambassador of the newly formed state of Telangana'}, dependency relations are as follows. dobj(made-4, Which-1), nsubjpass(m ade-4, sportsperson-2), auxpass(made-4, was-3), root (ROOT-0, made-4), det(ambassdor-7, the-5), compound(ambassdor-7, brand-6), dobj(made-4, ambassdor-7), case(stat e-11, of-8), advmod(formed-10, newly-9), amod(state-11, formed-10), nmod:of(amb assdor-7, state-11), case(Telangana-13, of-12), nmod:of(state-11, Telangana-13). The structural features are designed using the dependency principals of the proposed algorithm that is shown in Algorithm 3.4. The root word and its siblings are expanded to measure the design principles.

The TD includes many relations which are considered as structural features: For instance, in the sentence \textit{`Indian diplomat Devyani Khobragade posted where, when she was arrested in a visa case in 2013'}, The following relations under the TD representation are obtained: \\ $TD_1:$ \textit{amod}(khobragade-4, Indian-1) \\
$TD_2:$ \textit{det}(case-15, a-13) \\
$TD_3:$ \textit{compound}(case-15, visa-14) \\
$TD_4:$ \textit{nmod}(arrested-11, case-15) \\ The algorithm extracts four structural relations numeric modifier relation between \textit{`arrested'} and \textit{`case'}. Algorithm also provides an apposition relation between \textit{`posted'} and \textit{`arrested'}. The relation between these words represent the best possible link available in the text.  For example, the adjectival modifier gleeful in the sentence, relation of verb to have textual dependecy is shown: \\ $S_tF_1:$ \textit{dep}(posted-5, where-6) \\
$S_tF_2:$ \textit{advmod}(arrested-11, when-8) \\
$S_tF_3:$ \textit{advcl}(posted-5, arrested-11) \\
$S_tF_4:$ \textit{nmod}(arrested-11, case-15)

The connection between these outcomes shows that SF proposes a wider set of dependencies, catching relation distance which can contribute to evidence gathering and question alignment. The parallel structural representations help in linking two words which can not be linked otherwise, and this is the reason for choosing NP words as root.

The TD scheme offers the option prepositiona dependencies involvement. In the example \textit{`Name the  first deaf-blind person who  receive a bachelor of arts degree?'} instead of having two relations \textit{case}(degree-12, of-10) and \textit{dobj}(receive-7, bachelor-9) or \textit{nmod}(bachelor-9, degree-12) and \textit{acl:relcl}(person-5, receive-7), SF gives a relation between the pharses: \textit{case}(degree-12, person-5). These links are used later in this work in EG \& NLQA. Some more useful structural extractions such as, e.g. \textit{`Which sport uses these terms reverse swing and reverse sweep?'}. TD gives direct links between \textit{`swing'} and \textit{`swap'} for (\textit{dobj}),\\ $TD_5: $\textit{dobj} such as (reverse-6, swing-7) \\
$TD_6: $\textit{dobj} such as (reverse-9, sweep-10) \\
$SF_5: $ (reverse, sweep, \textit{dobj})

The information in $SF_5$ is not apparent in the TD which follows \textit{dobj} in a similar way, have relations with three parameters such as, ($SF_i$, $SF_j$, \textit{TD}). 

\begin{figure}[ht!]
	\centering
	\includegraphics[width=0.48\textwidth]{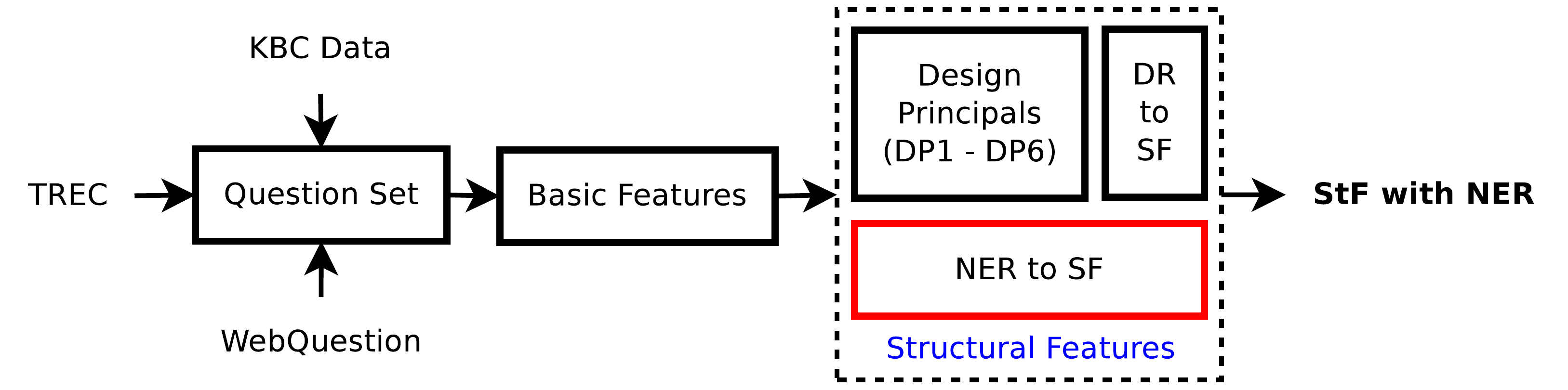}
	\caption{Adding named entities to structural features}\label{fig:39}
\end{figure}

SF representation is enhanced with the addition of named entities, for the sentence in Figure \ref{fig:39}. In Figure \ref{fig:39} structural features are extracted from design principles, dependency rules and named entities which give an outcome as structural features with NER. The information available for the word \textit{Telangana} in the SF scheme: $SF_{ne}:$ (Telangana∼5, location).
The structural information becomes valuable with the use of named entities and, SF provides the root to relate the words from the named entities. The Structural feature extraction algorithm using DR \& DP extraction rules is shown below in Algorithm 4.

\begin{algorithm}
	\caption{\textbf{Algorithm 4:} Structural feature extraction algorithm \& weight calculation}
	\label{algo:34}
	\textbf{INPUT:} Question set (Q) \\
	\textbf{OUTPUT:} $Sy_{fv}$ $\in$ Structural feature vector from Q\\
	\textbf{Variables Used:}\\
	 $(Q_f, V) \in (Question Feature, Feature Value)$  
	\begin{algorithmic} [1]
		\FOR {questions in dataset $Q$} 
		\IF{$isTerminal = `leaf'$}
		\STATE backtrack tree
		\ELSE
		\STATE expand\_root procedure
		\ENDIF
		\ENDFOR
		\STATE \textbf{procedure} expand\_root
		\IF {$root$ has child nodes}
		\FOR {childs in \textbf{tree} $T$} 
		\IF{$isTerminal \neq  `NP'$}
		\STATE backtrack \textbf{tree}
		\ELSE
		\STATE head\_child $\leftarrow$ apply\_rules from $DP$ (Rule 1 to 6)
		\STATE $Weight_{DP} \leftarrow Weight_{DP} + 1$
		\STATE head\_child $\leftarrow$ apply\_rules from $DR$
		\STATE $Weight_{DR} \leftarrow Weight_{DR} + 1$
		\STATE head\_child $\leftarrow$ apply\_NER
		\STATE $Weight_{NER} \leftarrow Weight_{NER} + 1$
		\ENDIF
		\ENDFOR
		\ENDIF
		
		\STATE $St_{fv} = Weight_{DP} + Weight_{DR} + Weight_{NER} $
		\STATE \textbf{Return} $ \rightarrow StructuralFeatureVector$ $(St_{fv})$ 
	\end{algorithmic}
\end{algorithm} 

\subsubsection{Feature Alignment with Individual Featuure Score}
It is important to calculate the individual feature value. These values are provided to a formula for final feature score. The formula is tested over 100 KBC questions having at least two similar questions. Extraction algorithms for all features have been discussed in earlier sections. Document for example feature score has a passage \textit{Indian tennis star Sania Mirza was today appointed 'Brand Ambassador' of Telangana}.

\textit{\textbf{Lexical Score-}} It is shown here that how relevant lexical features are extracted from a question. For an example from KBC dataset, \textit{`Which sportswoman was made the brand ambassador of the newly formed state of Telangana?'} $\in$ \textbf{\large Q} and, Indian tennis star Sania Mirza was today appointed \textit{`Brand Ambassador'} of Telangana? $\in$ \textbf{\large D}. Lexical features are extraction here and individual feature score is calculated from these is shown in Table \ref{table:ffs}. Equation \ref{eq:ra1} is showing the value of $R^2$ for Unigram features. Similarly Equation \ref{eq:ra2}, \ref{eq:ra3}, \ref{eq:ra4}, \ref{eq:ra5}, and \ref{eq:ra6} are showing the respective feature line and value of $R^2$ of Bigram, Trigram, Wh-word, Word Shape and Question length feature respectively.

\begin{table}
	\caption {\footnotesize{Calculation of average feature score to get the final feature form score}}
	\begin{center}
		\scalebox{0.80}{
			\begin{tabular}{|c|c|l|l|}
				\hline
				
				\multirow{6}{*}{\rotatebox{90}{Lexical}} & 1 &  $U_n$ & $U_n Score = \frac{TF}{SQ_{tf}}
				= \frac{5}{12}
				= 0.417$ \\
				
				& 2 &  $B_i$ & $B_i Score = \frac{TF}{SQ_{tf}}
				= \frac{3}{12}
				= 0.25$ \\
				
				& 3 &  $T_i$ & $T_i Score = \frac{TF}{SQ_{tf}}
				= \frac{5}{12}
				= 0.417$ \\
				
				& 4 &  $W_w$ & $W_w Score = \frac{W_wW}{SQ_{W_ww}}
				= \frac{1}{12}
				= 0.083$ \\
				
				& 5 &  $W_s$ & $W_s Score = \frac{WS}{WS_{SQ}}
				= \frac{4}{12}
				= 0.33$ \\
				
				& 6 & $Q_l$ & $Q_l Score = \frac{Q^nL}{SQ^nL}
				= \frac{12}{13}
				= 0.923$ \\ 
				\hline
				&&&\\
				
				& & \multicolumn{1}{c|}{\LARGE{$L_e$}} &    {\textbf{Average = $\frac{\sum L_e Score}{\sum No. of L_e}$ = $\frac{2.24}{6}$ = 0.403}    }  \\
				&&&\\
				\hline
				
				\multirow{5}{*}{\rotatebox{90}{Syntactic}} & 1 &  $T_u$ & $T_u Score = \frac{TU}{SQ_{tu}}
				= \frac{12}{13}
				= 0.33$ \\
				
				& 2 & $P_t$ & $P_t Score = \frac{PT}{SQ_{pt}}
				= \frac{5}{7}
				= 0.417$ \\
				
				& 3 &  $H_w$ & $H_w Score = \frac{HW}{SQ_{hw}}
				= \frac{10}{12}*\frac{10}{13}
				= 0.638$ \\
				
				& 4 & $H_t$ & $H_t Score = \frac{HT}{SQ_{ht}}
				= \frac{20}{12}*\frac{10}{13}
				= 0.127$ \\
				
				& 5 & $F_w$ & $F_w Score = \frac{FW}{SQ_{fw}}
				= \frac{2}{12}
				= 0.166$ \\
				
				\hline
				&&&\\
				
				& & \multicolumn{1}{c|}{\LARGE{$S_y$}} &    {\textbf{Average = $\frac{\sum S_y Score}{\sum No. of S_y}$ = $\frac{1.678}{5}$ = 0.335}    }  \\
				&&&\\
				\hline
				
				\multirow{3}{*}{\rotatebox{90}{Semantic}} & 1 & $H_h$ & $H_h Score = \frac{HW_h}{SQ_{hWh}}
				= \frac{2}{3}
				= 0.667$ \\
				
				& 2 & $N_e$ & $N_e Score = \frac{NE}{SQ_{ne}}
				= \frac{3}{5}
				= 0.60$ \\ 
				
				& 3 & $H_n$ & $H_n Score = \frac{H_{ne}}{SQ_{hne}}
				= \frac{2}{5}
				= 0.40$ \\
				\hline
				&&&\\
				
				& & \multicolumn{1}{c|}{\LARGE{$S_e$}} &    {\textbf{Average = $\frac{\sum S_e Score}{\sum No. of S_e}$ = $\frac{1.667}{3}$ = 0.556}    }  \\
				&&&\\
				\hline
				& \multicolumn{3}{c|}{} \\
				& \multicolumn{3}{c|}{{$ FF_b Score= \sum_{i = 1}^n [log (L_e{\times}S_y{\times}S_e)] = [log (0.403{\times}0.335{\times}0.556)]  $}} \\
				& \multicolumn{3}{c|}{} \\
				
				& \multicolumn{3}{c|}{} \\
				& \multicolumn{3}{c|}{\small{$ = [log (L_e) + log (S_y) + log (S_e)] = 0.264 + 0.252 + 0.274 = \large\textbf{0.79}  $}} \\
				& \multicolumn{3}{c|}{} \\
				\hline
				
		\end{tabular}}
	\end{center}
	\label{table:ffs} 
\end{table}

\textbf{\textit{1) Unigrams ($U_n$)- }} Unigrams of the questions are tagged as, {(Which, 1) (sportswoman, 2) (was, 3) (made, 4) (the, 5) (brand, 6) (ambassador, 7) (of, 8) (the, 9) (newly, 10) (formed, 11) (state, 12) (of, 13) (Telangana, 14)}. Refer table \ref{table:ffs} to see the feature score calculation of Unigram. $U_n$ feature regression line is shown in Equation \ref{eq:ra1}.

\begin{equation}
U_n = Y_1 = 0.001x_1 + 0.529 - {(R_1^2 = 0.18)}
\label{eq:ra1}
\end{equation}

\textbf{{2) Bigrams ($B_i$)- }} Bigrams of the questions are tagged as, {(Which-sportswoman, 1) (sportswoman-was, 2) (was-made, 3) (made-the, 4) (the-brand, 5) (brand-ambassador, 6) (ambassador-of, 7) (of-the, 8) (the-newly, 9) (newly-formed, 10) (formed-state, 11) (state-of, 12) (of-Telangana, 13)}. Refer table \ref{table:ffs} to see the feature score calculation of $B_i$ and the regression line is shown in Equation \ref{eq:ra2}.

\begin{equation}
B_i = Y_2 = 0.041x_2 + 0.639 - {(R_2^2 = 0.13)}
\label{eq:ra2}
\end{equation}

\textbf{\textit{3) Trigrams ($T_r$)- }} Trigrams of the questions are tagged as, {(Which-sportswoman-was, 1) (sportswoman-was-made, 2) (was-made-the, 3) (made-the-brand, 4) (the-brand-ambassador, 5) (brand-ambassador-of, 6) (ambassador-of-the, 7) (of-the-newly, 8) (the-newly-formed, 9) (newly-formed-state, 10) (formed-state-of, 11) (state-of-Telangana, 12)}. Refer table \ref{table:ffs} to see the feature score calculation of Trigram. $T_r$ feature regression line is shown in Equation \ref{eq:ra3}.

\begin{equation}
T_r = Y_3 = 0.061x_3 + 0.739 - {(R_3^2 = 0.23)}
\label{eq:ra3}
\end{equation}

\textbf{\textit{3) Wh-word ($W_w$), Word Shape  ($W_s$) and Q-Length   ($Q_l$)- }} $W_w$ word of the example is \textit{which}, $W_s$ feature provide (UPPERCASE, 2), $Q_l = 13$. Refer table \ref{table:ffs} to see the feature score calculation of Wh-word, Word Shape and Question Length. The feature regression line of $W_w, W_s$ and $Q_l$ is shown in Equation \ref{eq:ra4}, \ref{eq:ra5}, and \ref{eq:ra4}.

\begin{equation}
W_w = Y_4 = 0.001x_4 + 0.529 - {(R_4^2 = 0.38)}
\label{eq:ra4}
\end{equation}

\begin{equation}
W_s = Y_5 = 0.063x_5 + 0.542 - {(R_5^2 = 0.32)}
\label{eq:ra5}
\end{equation}

\begin{equation}
Q_l = Y_6 = 0.052x_6 + 0.461 - {(R_6^2 = 0.42)}
\label{eq:ra6}
\end{equation}

\subsection{Multiple Regression Analysis on Features}
\begin{figure}[ht!]
	\centering
	\includegraphics[width=0.48\textwidth]{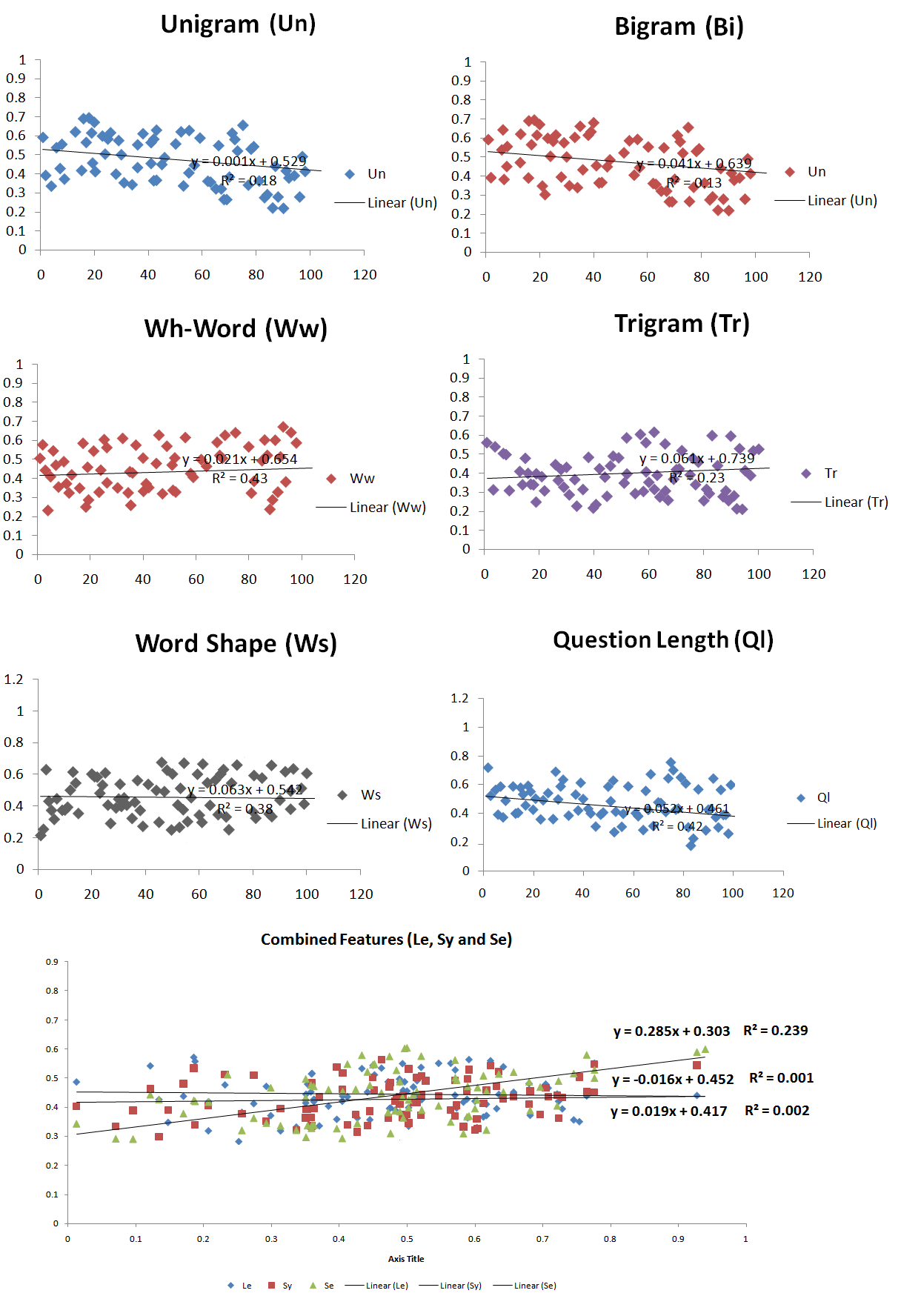}
	\caption{Regression line of faetures to calculate the QFF, DFF formuula}
	\label{fig:lfs}
\end{figure}

To calculate the final score of basic and proposed features, the formula is obtained from Multiple Regression (MR) techniques on feature scores. In MR the value of \textbf{$R^2$}, also known as the coefficient of determination is generally used statistic to estimate model fit. \textbf{$R^2$} is 1 minus the ratio of residual variability. While the variability of the residual values nearby the regression line corresponding to the overall variability is small, regression equation is well fitted. For example, if there is no link between the X and Y variables, then the ratio of the residual variability (Y variable) to the initial variance is equal to 1.0. Then \textbf{$R^2$} would be 0. If X and Y are perfectly linked then the ratio of variance would be 0.0, making \textbf{$R^2$} = 1. In most cases, the ratio and \textbf{$R^2$} will be between these 0.0 and 1.0. If \textbf{$R^2$}  is 0.2 then we understand that the variability of the Y values nearby the regression line is 1-0.2 times the original variance; in other words, we have explained 20\% of the original variability and left with 80\% residual variability. The equation \ref{eq:ra} is used for all the feature and tested on MR of available features. The regression line to of the individual feature to calculate QFF and DFF is shown in Figure \ref{fig:lfs}.

\begin{equation}
Y = QFF Score = a + b_1*L_e + b_2*S_y + b_3*S_e
\label{eq:ra}
\end{equation}

\subsubsection{Intermediate QFF Score} 
The logical form is used to query a knowledge base. Intermediate QFF generated in this work is not bothered about querying any KB, but it is used to represent the question to its QFF weight and then to map it with another QFF weights. These QFF weights are the QFF scores calculated from the formula shown in equation \ref{eq:formula} (generated from RA of features). In the equation \ref{eq:formula2} $L_e$ represents the lexical features, $S_y$ represents the syntactic features, $S_e$ represents the semantic features and $S_t$ represents the structural features. In the equation \ref{eq:ra} putting the value of $a=0$ and as the value is used for the alignments the effect of coefficient can be ignored once and treated as $b_1 = b_2 = b3 = 1$. QFF score is shown in equation \ref{eq:formula}. 

\begin{equation}
FF Score (QFF/DFF) = L_e + S_y + S_e \times S_t
\label{eq:formula}
\end{equation}

\begin{figure}[ht!]
	\centering
	\includegraphics[width=0.48\textwidth]{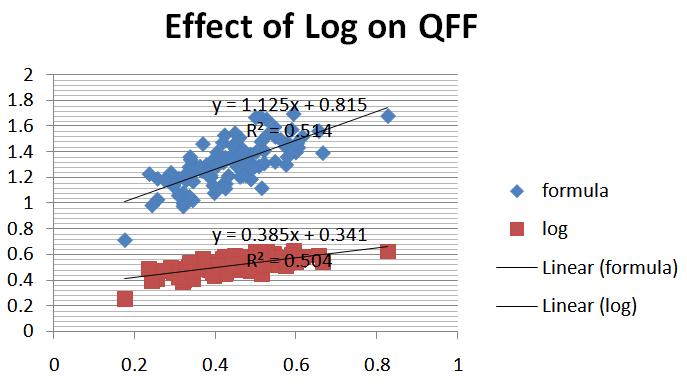}
	\caption{Diagram showing the effect of log}
	\label{fig:fig1}
\end{figure}

\begin{equation}
FF Score = (log L_e + log S_y + log S_e) \times S_t
\label{eq:formula2}
\end{equation}

QFF and also a (Document Feature Score) DFF score can be compared to question and document as there are about each other. One can also use multiple regression coefficients to compare QFF and DFF. In this work, the complete dataset has questions paired with options and answers and documents having the answer-evidence are ranked. The equation is merely showing that all feature are contributing equally to calculate QFF and QFF score. Equation \ref{eq:formula3} calculates the FFScore of each question in dataset.

\begin{equation}
FF Score = \sum_{i = 1}^n [log (L_e{\times}S_y{\times}S_e)]{\times}S_t
\label{eq:formula3}
\end{equation}

\subsection{Proposed Feature Relevance Technique}
Individual features are tested on the different dataset to get the final answer (features are used in QA). The feature which is contributing in attaining the highest accuracy by QA system is marked as the most relevant feature. The accuracy of answer correctness after including these individual features is shown in Table 3.

\begin{table}[!htb]
	\caption {\footnotesize{Basic and proposed features with their relevance in QA}}
	
	\begin{center}
		
			\begin{tabular}{|l|c|c|c|c|}
				\hline
				& \multicolumn{3}{c|}{\textbf{Correct Answers (\%)}} & \\
				\hline
				\textbf{Basic Features} &    {WebQ} &    {TREC} & {KBC} &      \textbf{Relevance} (\textbf{Fr: 1-5})   \\
				\hline
				Unigrams & 61 & 63 & 67 &    \textbf{4} \\ \hline
				Bigrams & 82 & 79 & 88 &    \textbf{5} \\ \hline
				Trigrams & 58 & 55 & 52 &    \textbf{3} \\ \hline
				Wh-word & 48 & 35 & 32 &    \textbf{3} \\ \hline
				Word Shape & 51 & 43 & 48 &    \textbf{3} \\ \hline
				Question Length & 28 & 23 & 19 &    \textbf{2} \\ \hline
				Tagged Unigram & 43 & 42 & 46 &    \textbf{3} \\ \hline
				POS tags & 46 & 51 & 56 &    \textbf{3} \\ \hline
				Headword & 87 & 88 & 91 &    \textbf{5} \\ \hline
				Headword Tag & 62 & 58 & 52 &    \textbf{4} \\ \hline
				Focus Word & 76 & 72 & 80 &    \textbf{4} \\ \hline
				HW Hypernyms & 66 & 54 & 63 &    \textbf{4} \\ \hline
				Named Entity & 83 & 82 & 77 &    \textbf{5} \\ \hline
				Headword NE & 57 & 52 & 49 &    \textbf{3} \\ \hline
				\multicolumn{5}{|l|}{\textbf{Proposed features (structural features $S_t$)}}  \\ \hline
				$S_t$ with DP& 56 & 61  & 65 &    \textbf{4} \\ \hline
				$S_t$ with DR& 67 & 68 & 72 &    \textbf{4} \\ \hline
				$S_t$ with NER& 92 & 88 & 91 &    \textbf{4} \\
				\hline
		\end{tabular}
	\end{center}
	
	\label{table:32} 
\end{table}

The feature relevance is calculated by the Equation \ref{eq:fr}, where $\sum QC_i$ is the sum of correctly answered questions (i $\in$ KBC, WebQuestions and TREC) and $\sum Q_T$ is the total number of questions.  
\begin{equation}
\begin{aligned}\label{eq:fr}
F_r = \frac{1}{2} \times \frac{\sum QC_i}{\sum Q_T}
\end{aligned}
\end{equation}
The relevance score of the features ($F_r$) is useful where feature vector is redundant and we need to reduce the space. In such situations the features with low relevance score can be removed from the feature vector. The feature selection techniques (Agarwal et al. 2014) are also used for selecting the features with relevant information.  

\section{Dataset and Result Analysis}
\subsection{Dataset Used}
To measure the correctness of the proposed features and their extraction algorithms, the publicly available KBC dataset is used. This dataset consists of open-domain questions with option and answers. For accurate and more stable experiments, TREC and WebQuestions datasets are also used that consist the relevant documents. 

\subsection{Performance Metrices}
The performance of the feature extraction algorithms on KBC dataset and other datasets is measured by the total number of questions accurately answered by each features and by the combination of features.

\begin{itemize}
	\item[] \textbf{Correct Answers (CA)}: It belongs to the number of correct answers provided by a particular feature. 
	\item[] \textbf{Incorrect Answers (IA)}: It belongs to the number of incorrect answers provided by a particular feature. 
	\item[] \textbf{Correct Documents (CD)}: It belongs to the number of correct documents selected by a particular feature. 
	\item[] \textbf{Incorrect Documents (ID)}: It belongs to the number of incorrect documents selected by a particular feature. 
\end{itemize}

The feature accuracy is employed for estimating the performance of basic and proposed features in QA. The precision of features is the division of the total questions that are correctly expressed by the features and the total number of documents that are to be expressed (it is the summation of TP and FP) as given in Equation \ref{eq:34}. 
\begin{equation}
Precision_{features} = \frac{CA}{CA + CD}
\label{eq:33}
\end{equation}
The recall is the division of the total number of correctly expressed question or documents to the total number of question or documents that are to be expressed (it is the sum of TP and FN) as given in Equation \ref{eq:35}.
\begin{equation}
Recall_{features} = \frac{CA}{CA + ID}
\label{eq:34}
\end{equation}
F-measure is the aggregate of $Precision_{features}$ and $Recall_{features}$ is given by Equation \ref{eq:35}.
\begin{equation}
Answer Extraction_{accuracy} = \frac{2 \times Precision \times Recall}{Precision + Recall}
\label{eq:35}
\end{equation}
In this analysis, accuracy of answer extraction ($A_cAE$) (also termed as F-measure) is done to report the performance of feature representation for QA systems and later NLQA and EG algorithms.

\subsection{Results and Discussions}
Ten-fold cross validation method is employed to estimate the accuracy of the proposed methods. The question data and documents are split randomly into 90\% training and 10\% testing. Wh-words (who, where, when) give an idea of expected answer type of the question, that is why it is important to handle such wh-words in QA. In the experiments, a simple approach is adopted such as to find the Expected Answer Type in the document. For example, \textit{`which is the largest city of the world'}, the \textit{`city'} is the EAT for this question and the document is searched for all the named entities with NE tag \textit{`Location'}. The weighting formula for individual features has been discussed in Algorithm 1, 2 and 3 are used to calculate the weight of the basic features and proposed features. 

\subsubsection{Determination of Prominent Features}
In the experiments it can be observed that bigrams ($B_i$) features are better than any other features on datasets as shown in Table 4. The bigram feature set provides the accuracy 69\% as compared to 62\%, 58\%, and 52\% for unigrams, trigrams and word-shape feature respectively on KBC dataset as shown in Table 4. The probable causes for this can be explained as follows. Unigram feature set contains lots of irrelevant features, which depreciates the accuracy of answer extraction. Also, trigrams feature set scattered than bigrams which demote the accuracy. Word-shape features are not valuable for the question and important mainly for document analysis. Word-shape feature set contains less information that is not enough for answer extraction in QA. Hence these perform worst when used separately. The dependency features are present in both question and document, resulting in more particular features. That is why these features are used to design structural features and contribute much for QA. These standard features are more useful for QA. 

Table \ref{table:33} present the accuracy of answer extraction ($A_c{AE}$) for all the basic features and Table \ref{table:34} present the accuracy of answer extraction for all the proposed features. The accuracy of bigram features is considered as the baseline for this experiments.

\begin{table*}[!htb]
	\caption {\footnotesize{Accuracy of Answer Extraction ($A_cAE$) of basic features on KBC dataset}}
	\begin{center}
		
			\begin{tabular}{|c|c|l|c|c|c|}
				\hline
				\multicolumn{6}{|l|}{\textbf{Basic Features on KBC Dataset}}    \\
				\hline
				& & & $A_cAE$ (\%) & & $A_cAE$ (\%)   \\
				\hline
				\multirow{7}{*}{\rotatebox{90}{Lexical}} & 1 & Unigram $U_n$ & 62 & $U_n + B_i$ & 84 \\
				
				& 2 & Bigram $B_i$ & 69 & $B_i + U_n$ & 84 \\
				
				& 3 & Trigram $T_i$ & 58 & $T_r + U_n$ & 66 \\
				
				& 4 & Wh-word $W_w$ & 38 & $W_w + B_i$ & 72 \\
				
				& 5 & Word Shape $W_s$ & 52 & $W_s + B_i$ & 67 \\
				
				& 6 & $Q^n$ Length $Q_l$ & 18 & $Q_l + B_i$ & 64 \\
				\cline{2-6}
				
				& \textbf{(a)} & \multicolumn{1}{c|}{{$Le_{avg}$}} & 49.5  & \textbf{$U_n + B_i$} & 84 \\
				
				\hline
				
				& \textbf{(b)} &\multicolumn{1}{c|}{$Le_{max}-Le_{avg}$}& +19.5 & $Le^*_{max}-Le^*_{avg}$  & +34.5  \\
				\hline
				
				\multirow{6}{*}{\rotatebox{90}{Syntactic}} & 1 & Tagged $U_n$ $T_u$ & 45 & $T_u + H_w$ & 78 \\
				
				& 2 & POS Tag $P_t$ & 52 & $P_t + H_t$ & 58 \\
				
				& 3 & Headword $H_w$ & 62 & $H_w + T_u$ & 78 \\
				
				& 4 & $H_w$ Tag $H_t$ & 53 & $H_t + P_t$ & 70 \\
				
				& 5 & Focus Word $F_w$ & 61 & $F_w + T_u$ & 72 \\
				
				\cline{2-6}
				& \textbf{(c)} & \multicolumn{1}{c|}{{$Sy_{avg}$}} & 54.6  & \textbf{$U_n + B_i$} & 71.2 \\
				
				\hline
				
				& \textbf{(d)} &\multicolumn{1}{c|}{$Sy_{max}-Sy_{avg}$}& +7.4 & $Sy^*_{avg}-Sy^*_{max}$  & +6.8  \\
				\hline
				
				& \textbf{(e)}  & \multicolumn{1}{c|}{{$Le_{avg} + Sy_{avg}$}} & 52.1 & \textbf{$U_n + B_i + H_w + T_u$} & 83.6 \\
				\hline
				
				& \textbf{(f)} &\multicolumn{1}{c|}{$Le_{max}-Sy_{avg}$}& +14.4 & $Le^*_{max}-Sy^*_{avg}$  & +12.8  \\
				\hline

				\multirow{4}{*}{\rotatebox{90}{Semantic}} & 1 & $H_w$ Hypernym $H_h$ & 44 & $H_h + N_e$ & 78 \\
				
				& 2 & Named Entity $N_e$ & 65 & $N_e + H_h$ & 78 \\
				
				& 3 & $H_w$ NE $H_n$ & 62 & $H_n + N_e$ & 67 \\
				\cline{2-6}
				& \textbf{(g)}  & \multicolumn{1}{c|}{{$Se_{avg}$}} & 57 & \textbf{$H_h + N_e$} & 74.3 \\
				\hline
				
				& \textbf{(h)} &\multicolumn{1}{c|}{$Se_{max}-Sy_{avg}$}& +8 & $Se^*_{avg}-Se^*_{max}$  & +3.7  \\
				\hline
				
				& \textbf{(i)} &\multicolumn{1}{c|}{$Se_{max}-Le_{avg}$}& +15.5 & $Se^*_{avg}-Le^*_{max}$  & -27  \\
				\hline

				&  \textbf{(j)} & \multicolumn{1}{c|}{{$L_e + S_y + S_e$}} & 53.7 & \textbf{$U_n + B_i + H_w + T_u$} & 89.6 \\
				\hline
				
				& \textbf{(k)} &\multicolumn{1}{c|}{$Le_{max}-Se_{avg}$}& +12 & $Le^*_{avg}-Se^*_{max}$  & +32  \\
				\hline
				
				& \textbf{(l)} &\multicolumn{1}{c|}{$Sy_{max}-Se_{avg}$}& +5 & $Sy^*_{avg}-Se^*_{max}$  & +35  \\
				\hline
		\end{tabular}
	\end{center}
	\label{table:33} 
\end{table*}

\begin{table}[!htb]
	\caption {\footnotesize{Accuracy of Answer Extraction ($A_cAE$) of proposed features on KBC dataset}}
	\begin{center}
		\scalebox{0.80}{
			\begin{tabular}{|c|c|c|c|c|c|}
				\hline
				\multicolumn{6}{|l|}{\textbf{Proposed Features on KBC Dataset}}    \\
				\hline
				& & & $A_cAE$ (\%) & & $A_cAE$ (\%)   \\
				\hline
				
				\multirow{4}{*}{\rotatebox{90}{Structural}} & 1 & $S_tF$ with DR  $St_{dr}$ & 64 & $St_{dr} + St_{ner}$ & 78 \\
				
				& 2 & $S_tF$ with DP $St_{dp}$ & 68 & $St_{dp} + St_{ner}$ & 82 \\ 
				
				& 3 & $S_tF$ with NER $St_{ner}$ & 58 & $St_{dr} + St_{dr}$ & 78 \\
				
				\cline{2-6}
				
				& \textbf{(m)}  & \multicolumn{1}{c|}{{$St_{avg}$}} & 63.3  & \textbf{$St_{dp} + St_{ner}$} & 82 \\
				
				\hline
				
				& \textbf{(n)} &\multicolumn{1}{c|}{$Le_{max}-St_{avg}$}& 5.7 & $Le^*_{avg}-St^*_{max}$  & +32.5  \\
				\hline
				
				& \textbf{(o)} &\multicolumn{1}{c|}{$Sy_{max}-St_{avg}$}& 1.3 & $Sy^*_{avg}-St^*_{max}$  & +27.4  \\
				\hline
				
				&  \textbf{(p)} &\multicolumn{1}{c|}{$Se_{max}-St_{avg}$}&6.3 & $Se^*_{avg}-St^*_{max}$  & +25  \\
				\hline
		\end{tabular}}
	\end{center}
	\label{table:34} 
\end{table}

The proposed feature sets features with the addition of DP, DR, and NER increases the accuracy for datasets (WebQuestions, TREC, and KBC). For example, $St_{dr}$ features increased the accuracy from 64\% to 78\% (+14\%) with addition of $St_{ner}$ on KBC dataset. The proposed $St_{dp}$ features also increased the accuracy from 68\% to 82\% (+14\%) with addition of $St_{ner}$ on KBC dataset. It is because adding NER to structural features improve the accuracy by dropping unnecessary and unrelated features. Structural features with design principals $St_{dp}$ in addition to NER  attain the accuracy of 82\% as compared to its comparable combined basic feature set ($U_n + B_i + H_w + T_u$) i.e. 89.6\%. It is 7.4\% more than the proposed structural features but still meaningful. Therefore, structural features are very relevant and selective features. It is clear that in basic features, bigram features performed well than others, while used independently. Whereas, proposed features are concerned dependency based structural features are more prominent than basic features (including bigram features), and the accuracy over datasets is presented in Table \ref{table:35}.

			\begin{table*}
	\caption {\footnotesize{Basic features on each question datsets and comparisions after adding structural features}}
	\begin{center}
		
			\begin{tabular}{|c|c|l|c|c|c|c|}
				\hline
				\multicolumn{7}{|l|}{\textbf{Without structural feature ($S_tf$)}}     \\
				\hline
				&   &  & \multicolumn{4}{c|}{\textbf{Question Dataset (No. of question taken)}}   \\
				\hline
				
				&  &  & WebQ (W) & TREC (T) & KBC (K)  & (W + T + K)     \\
				&  &  & (5500) & (2000) & (500)  & (300 + 300 + 300)     \\
				\hline
				
				\multirow{7}{*}{\rotatebox{90}{Lexical}} & 1 &  $U_n$ & 63.2 & 60.1   & 67.3   & 65.2     \\  \cline{2-7}
				
				& 2 &  $B_i$ & 68.1  & 55.3   & 62.6    & 66.1     \\  \cline{2-7}
				
				& 3 &  $T_i$ & 59.2  & 56.3   & 62.4   & 58.3   \\  \cline{2-7}
				
				& 4 &  $W_w$ & 38.3  & 53.6   & 58.4   & 61.2    \\  \cline{2-7}
				
				& 5 &  $W_s$ & 52.1  & 56.8   & 64.7   & 67.1    \\  \cline{2-7}
				
				& 6 & $Q_l$ & 16.1  & 19.4   & 21.2   & 23.2     \\  
				\cline{2-7}
				&  & \textbf{$Le_{avg}$} & \textbf{49.5}  & \textbf{50.25}   & \textbf{56.03}   & \textbf{56.85}   \\  \cline{2-7}
				\cline{1-7}
				
				\multirow{6}{*}{\rotatebox{90}{Syntactic}} & 1 & $T_u$ & 43.2  & 55.8     & 57.5   & 53.6    \\   \cline{2-7}
				
				& 2 & $P_t$ & 50.8  & 52.1   & 61.9   & 64.3   \\  \cline{2-7}
				
				& 3 & $H_w$ & 60.8  & 86.5   & 71.3   & 70.6   \\  \cline{2-7}
				
				& 4 &  $H_t$ & 57.2  & 58.9   & 61.4   & 66.3   \\  \cline{2-7}
				
				& 5 &  $F_w$ & 53.8  & 55.9   & 59.6   & 55.4  \\ \cline{2-7}
				
				&  & \textbf{$Sy_{avg}$} & \textbf{53.16}  & \textbf{61.84}    & \textbf{62.34}   & \textbf{62.02}   \\
				\cline{1-7}
				\multirow{4}{*}{\rotatebox{90}{Semantic}} & 1 &  $H_h$ & 44.5  & 55.3   & 75.8     & 74.7   \\  \cline{2-7} 
				
				& 2 &  $N_e$ & 65.1  & 55.8   & 73.9     & 71.6   \\  \cline{2-7} 
				
				& 3 & $H_n$ & 60.5  & 52.4   & 68.3     & 72.9   \\   \cline{2-7}
				
				&  & \textbf{$Se_{avg}$} & \textbf{56.7}  & \textbf{54.5}   & \textbf{72.67}     & \textbf{73.06}   \\ 
				\cline{1-7}
				
				\multicolumn{7}{|l|}{\textbf{With structural feature ($S_tf$)}}   \\
				\cline{1-7}
				
				\multirow{7}{*}{\rotatebox{90}{Lexical}} & 1 &  $U_n$  & 68.2 (+5)  & 62.4  (+2.3) & 72.7 (+5.4)   & 71.7 (+6.5)  \\   \cline{2-7}
				
				& 2 &  $B_i$ & 69.8 (+1.7)  & 58.9 (+3.6)  & 71.1 (+8.5)    & 73.3 (+7.2)  \\  \cline{2-7}
				
				& 3 &  $T_i$ & 60.1 (+0.9)   & 57.7  (+1.4) & 63.2  (+0.8)    & 68.9 (+10.6)  \\  \cline{2-7}
				
				& 4 &  $W_w$ & 41.4 (+3.1)   & 55.5  (+1.9) & 61.2  (+2.8)    & 72.6 (+11.4)  \\  \cline{2-7}
				
				& 5 &  $W_s$ & 55.3 (+3.2)   & 62.4  (+6.1) & 57.7  (-7.0)    & 68.3 (+1.2)  \\  \cline{2-7}
				
				& 6 & $Q_l$ & 15.3 (-0.8)   & 19.5  (+0.1) & 22.1  (+0.9)    & 23.1 (-0.1)  \\  \cline{2-7}
				
				&  & \textbf{$Le_{avg}$} & \textbf{51.68 (+2.18)}   & \textbf{52.73 (+2.48)} & \textbf{58.0  (+1.97) } & \textbf{62.98 (+6.13) }  \\ \cline{1-7}

				\multirow{6}{*}{\rotatebox{90}{Syntactic}} & 1 & $T_u$ & 61.3 (+18.1)   & 55.9 (+0.1)  & 69.7 (+12.2)     & 75.9 (+18.4)  \\  \cline{2-7}
				
				& 2 & $P_t$ & 55.3 (+4.5)   & 56.7 (+4.6)  & 71.4 (+9.5)     & 74.3 (+10.0)  \\  \cline{2-7}
				
				& 3 & $H_w$ & 65.5 (+4.7)   & 86.9 (+0.4)  & 72.2  (+0.9)    & 66.9 (-3.7)  \\  \cline{2-7}
				
				& 4 &  $H_t$ & 58.1 (+0.9)   & 62.7 (+3.8)  & 72.3 (+10.9)     & 73.9 (+7.6)  \\  \cline{2-7}
				
				& 5 &  $F_w$ & 61.3 (+7.5)   & 56.4 (+0.5)  & 67.5  (+7.9)    & 71.3 (+15.9)  \\  \cline{2-7}
				
				&  & \textbf{$Sy_{avg}$} & \textbf{60.3 (+7.14)}   & \textbf{63.72 (+1.88)}  & \textbf{70.62 (+8.28)}     & \textbf{72.46 (+10.44)}  \\  \cline{2-7}
				
				\multirow{4}{*}{\rotatebox{90}{Semantic}} & 1 &  $H_h$ & 79.1 (+34.6)   & 69.4 (+14.1) & 76.2  (+0.4)   & 78.6 (+3.9)  \\   \cline{2-7}
				
				& 2 &  $N_e$ & 81.3  (+16.2)  & 78.3 (+22.5)  & 87.6 (+13.7)   & 82.4 (+10.8)  \\   \cline{2-7}
				
				& 3 & $H_n$ & 66.4 (+5.9)   & 71.2 (+18.8)  & 84.3 (+16.0)  & 79.9 (+7.0)   \\   \cline{2-7}
				
				&  & \textbf{$Se_{avg}$} & \textbf{75.6 (+18.9)}   & \textbf{72.9 (+18.4)}  & \textbf{82.7 (+10.03)}  & \textbf{80.3 (+7.24)}    \\ \cline{1-7}

		\end{tabular}
	\end{center}
	\label{table:35} 
\end{table*}        

The proposed structural features give better results than the basic bigram features with very fewer feature sizes. For example, question dataset after addition of structural features produced an accuracy of up to 82.7\% (+10.03\%) with 15 basic features for KBC adding the structural feature ($S_tf$) as shown in Table 3.4. Similarly, all the proposed features are constructed using dependency rules and performed better than similar basic features. For example, $Se_{avg}$ attained the accuracy of 73.06\%, whereas addition of structural features in these features produced an accuracy 80.3\% (+7.24) on a combination of WebQ, TREC and, KBC dataset as shown in Table 3.4. Structural feature set presents the accuracy of 82.7\% (+10.03\%) on KBC dataset. It is because by including the dependency rules, design principals, and NER relevant semantic information dependency features contain a large number of long distance relation capture. The proposed features produce an accuracy of 75.6\% with a maximum increase in the accuracy of +18.9 as shown in Table \ref{table:35}. 

The proposed features resolve the issue of hidden features while decreasing the feature space by combining the features with basic features and NER features. Structural features include the noun-phrase dependency distance as per design principals. Structural features are very useful to their ease of extraction, and these reduce the feature vector size significantly. It is followed in the experiments that if structural feature vector is very small then the performance is not well, and as feature vector size is increased the performance increases. It is due to the reason that as vector size is increased, the possibility of a grouping of the root words in the structure performs better. Experimental outcomes state that the proposed structural features with the addition to basic features perform better.

\section{Conclusion}
The performance of several basic features of relevant documents is examined on three datasets namely WebQuestions, TREC (8 and 9), and KBC. Apart from the basic features new structural features are proposed viz. structural features. The feature extraction algorithms for basic and proposed structural features is proposed. 

Proposed structural feature are combined with DP, with DR, and with NER. Further, the features have been assigned a relevance value which is calculated from the accuracy of an individual feature by their answer extraction accuracy on QA systems. It is also examined that addition of proposed structural features to the basic features improve the performance of answer extraction on QA systems. 

Furthermore, it is noticed that proposed structural features provide improved results for bigrams ($B_i$) features and prominent proposed structural with NER ($St_{ner}$) features provide excellent results than basic features. The accuracy of the question length features was near to the 20\% which is the minimum among all features. The main cause for this is the $Q_l$ only gives the idea of the question complexity. It is observed that when two basic features are combined their combination gives better results than the individual feature. The combination of bigrams with question length do not perform well for QA systems. 

All the features used in this work are useful to gather evidence and a indirect reference based approach for evidence gathering is proposed and combined with feature-based evidence gathering.


References are important to the reader; therefore, each citation must be complete and correct. If at all possible, references should be commonly available publications.


\begin{thebibliography}{99}

\bibitem{c1} Agarwal, B. and Mittal, N., \textit{Prominent feature extraction for review analysis: an empirical study}, Journal of Experimental \& Theoretical Artificial Intelligence, pp. 485-498, 2016. 

\bibitem{c2} Agarwal, B., and Mittal N., \textit{Semantic feature clustering for sentiment analysis of English reviews}, IETE Journal of Research, vol. 6, pp. 414-422, 2014. 

\bibitem{c3} Bishop, Lawrence C., and George Ziegler. ”Open ended question analysis system and method.” U.S. Patent No. 4,958,284. 18 Sep. 1990. 

\bibitem{c4} Brill E., Susan D., and Michele B., \textit{An analysis of the AskMSR question-answering system}, In Proceedings of the ACL-02 conference on Empirical methods in natural language processing, vol. 10, pp. 257-264, 2002 

\bibitem{c5} Bunescu R., Huang Y., \textit{Towards a general model of answer typing: Question focus identication}, In Proceedings of The 11th International Conference on Intelligent Text Processing and Computational Linguistics (CICLing), pp. 231-242, 2010. 

\bibitem{c6} Chowdhury G. G., \textit{Natural language processing}, In Annual review of information science and technology, vol. 1, pp. 51-89, 2003. 

\bibitem{c7} Corley C., and Mihalcea R., \textit{Measuring the semantic similarity of texts}, In Proceedings of the ACL workshop on empirical modeling of semantic equivalence and entailment, pp. 13-18, 2005. 

\bibitem{c8} Huang, Cheng-Hui, Jian Yin, and Fang Hou., \textit{A text similarity measurement combining word semantic information with TF-IDF method}, In Jisuanji Xuebao (Chinese Journal of Computers), pp. 856-864, 2011 

\bibitem{c9} Islam, Aminul, and Diana Inkpen., \textit{Semantic text similarity using corpus-based word similarity and string similarity}, In ACM Transactions on Knowledge Discovery from Data (TKDD) 2, pp. 10-19, 2008. 

\bibitem{c10} Jonathan B., Chou A., Roy F. and Liang P., \textit{Semantic Parsing on Freebase from Question-Answer Pairs}, In Proceedings of EMNLP, pp. 136-157, 2013. 

\bibitem{c11} Lally, Adam, John M. Prager, Michael C. McCord, Branimir K. Boguraev, Siddharth Patwardhan, James Fan, Paul Fodor, and Jennifer Chu-Carroll., \textit{Question analysis: How Watson reads a clue}, In IBM Journal of Research and Development, 2012. 

\bibitem{c12} Loni B., \textit{A survey of state-of-the-art methods on question classification}, In Literature Survey, Published on TU Delft Repository, 2011. 

\bibitem{c13} Miller G. A., \textit{WordNet: A Lexical Database for English}, In Communications of the ACM, vol. 38, pp. 39-41, 1995. 

\bibitem{c14} Mohler, M., and Rada irschmaMihalcea., \textit{Text-to-text semantic similarity for automatic short answer grading}, In Proceedings of the 12th Conference of the European Chapter of the Association for Computational Linguistics, pp. 567-575, 2009. 

\bibitem{c15} Pedoe W. T., \textit{True Knowledge: Open-Domain Question Answering Using Structured Knowledge and Inference}, In Association for the Advancement of Artificial Intelligence, vol. 31, pp. 122-130, 2014. 

\bibitem{c16} Severyn Aliaksei.,Moschitti Alessandro, \textit{Automatic feature engineering for answer selection and extraction}, In Proceedings of the 2013 Conference on Empirical Methods in Natural Language Processing, pp. 458-467, 2013. 

\bibitem{c17} Sharma L. K., Mittal N., \textit{An Algorithm to Calculate Semantic Distance among Indirect Question Referents}, International Bulletin of Mathematical Research, vol. 2, pp. 6-10, ISSN: 2394-7802, 2015. 

\bibitem{c18} Sharma LK, Mittal N. \textit{Prominent feature extraction for evidence gathering in question answering. Journal of Intelligent \& Fuzzy Systems}. 32(4):2923-32; Jan 2017.

\bibitem{c19} Singhal A. and Kaszkiel M, \textit{TREC-9}, In TREC at ATT, 2000 

\bibitem{c20} Voorhees, E.M., \textit{The TREC-8 Question Answering Track Report}, In TREC, vol. 99, pp. 77-82, 1999. 

\bibitem{c21} Wang, Z., Yan, S., Wang, H. and Huang, X., \textit{An overview of Microsoft deep QA system on Stanford WebQuestions benchmark}, Technical report, Microsoft Research, 2014. 

\bibitem{c22} Wei Xing, Croft Bruce. W, Mccallum Andrew, \textit{Table extraction for answer retrieval},In Information Retrieval, pp. 589-611, vol 9, 2006. 

\bibitem{c23} Yao X., Durme V. B., Clark P., \textit{Automatic Coupling of Answer Extraction and Information Retrieval}, In Proceedings of ACL short, Sofia, Bulgaria, pp. 109-116, 2013.






\end{thebibliography}
\end{document}